\documentclass[12pt]{article}
\pdfoutput=1

\usepackage{putex}

\usepackage{graphicx}
\usepackage{dsfont}
\usepackage{amsmath}
\usepackage{amssymb}
\usepackage{array}
\usepackage{mathdots}
\usepackage{epstopdf}
\usepackage{enumerate}
\usepackage{lipsum}
\usepackage{cite}
\usepackage{youngtab}
\usepackage{tensor}
\usepackage{braket}
\usepackage[normalem]{ulem}
\usepackage{slashed}
\usepackage[aligntableaux=center]{ytableau}
\usepackage[utf8]{inputenc}
\usepackage[
colorlinks=true,
linkcolor=blue,
urlcolor=blue,
filecolor=black,
citecolor=red,
]{hyperref}
\usepackage{braket}
\usepackage{mathtools}
\usepackage{rotating}
\usepackage{tabularx}
\normalsize\usepackage{booktabs}
\usepackage{caption}
\usepackage{subcaption}
\usepackage{bbm}
\newcolumntype{Y}{>{\centering\arraybackslash}X}
\newcolumntype{R}{>{\raggedright\arraybackslash}X}
\newcolumntype{C}[1]{>{\centering\arraybackslash}p{#1}}
\usepackage{xcolor}
\definecolor{LightCyan}{rgb}{0.7,1,1}
\definecolor{Gray}{gray}{0.9}
\usepackage{xspace}

\newcommand{\grp}[1]{\mathrm{#1}}

\newcommand{\grSU}{\grp{SU}}

\newcommand{\abs}[1]{\left\lvert #1 \right\rvert}

\newcommand {\be} {\begin {equation}}
\newcommand {\ee} {\end {equation}}

\newcommand {\bes} {\begin {equation*}}
\newcommand {\ees} {\end {equation*}}

\newcommand{\es}[2] {\begin{equation} \label{#1} \begin{split} #2 \end{split} \end{equation}}

\newcommand{\Z}{\mathbb{Z}}

\def\half{{\scriptstyle \frac 12}}

\newcommand{\beq}{\begin{equation}}
	\newcommand{\eeq}{\end{equation}}

\def\ie{\begin{equation}\begin{aligned}}
		\def\fe{\end{aligned}\end{equation}}

\def\<{\langle}
\def\>{\rangle}

\newcommand{\bkt}[2]{\langle#1|#2\rangle}

\def\pE{\mathcal{E}}

\def\beg{\begin{equation}\begin{gathered}}
		\def\eeg{\end{gathered}\end{equation}}

\def\bea{\begin{equation}\begin{aligned}}
		\def\eea{\end{aligned}\end{equation}}

\newcommand{\lmax}{\ell_\text{max}}
\newcommand{\gym}{g_\text{YM}}
\newcommand{\gDLCQ}{g_\text{DLCQ}}

\DeclareMathOperator{\spin}{\textsc{spin}}
\DeclareMathOperator{\adj}{\textsc{adj}}
\let\vector\relax
\DeclareMathOperator{\vector}{\textsc{vec}}
\DeclareMathOperator{\su}{\mathfrak{su}}
\DeclareMathOperator{\so}{\mathfrak{so}}

\numberwithin{equation}{section}

\begin{document}

\preprint{PUPT-2649}

\institution{PU}{Joseph Henry Laboratories, Princeton University, Princeton, NJ 08544, USA}
\institution{PCTS}{Princeton Center for Theoretical Science, Princeton University, Princeton, NJ 08544, USA}
\institution{IAS}{Institute for Advanced Study, Princeton, NJ 08540, USA}

\title{Lattice Hamiltonian for Adjoint QCD$_2$}

\authors{Ross Dempsey,\worksat{\PU} Igor R.~Klebanov,\worksat{\PU, \PCTS} Silviu S.~Pufu,\worksat{\PU, \PCTS, \IAS}\\[10pt] and Benjamin T. S\o gaard\worksat{\PU}}

\abstract{We introduce a Hamiltonian lattice model for the $(1+1)$-dimensional $\grSU(N_c)$ gauge theory coupled to one adjoint Majorana fermion of mass $m$.  The discretization of the continuum theory uses staggered Majorana fermions.  We analyze the symmetries of the lattice model and find lattice analogs of the anomalies of the corresponding continuum theory.   An important role is played by the lattice translation by one lattice site, which in the continuum limit involves a discrete axial transformation.
On a lattice with periodic boundary conditions, the Hilbert space breaks up into sectors labeled by the $N_c$-ality $p=0, \ldots N_c-1$.  Our symmetry analysis implies various exact degeneracies in the spectrum of the lattice model. In particular, it shows that, 
for $m=0$ and even $N_c$, the sectors $p$ and $p’$ are degenerate if $|p-p’| = N_c/2$. In the $N_c = 2$ case, we explicitly construct the action of the Hamiltonian on a basis of gauge-invariant states, and we perform both a strong coupling expansion and exact diagonalization for lattices of up to $12$ lattice sites. Upon extrapolation of these results, we find good agreement with the spectrum computed previously using discretized light-cone quantization. One of our new results is the first numerical calculation of the fermion bilinear condensate.}
\date{November 2023}

\maketitle

\tableofcontents

\section{Introduction}

The $(1+1)$-dimensional $\grSU(N_c)$ gauge theory coupled to one adjoint multiplet of Majorana fermions, sometimes referred to as adjoint QCD$_2$, is an interesting model of non-perturbative gauge dynamics.  As in other models where all the dynamical fields are in the adjoint representation of $\grSU(N_c)$, the Wilson loop in the fundamental representation serves as a precise criterion for confinement \cite{Wilson:1974sk}. 
Furthermore, similarly to the one-flavor Schwinger model \cite{Schwinger:1962tp}, this theory has only massive bound states even when the fermion is massless \cite{Dalley:1992yy,Kutasov:1993gq,Bhanot:1993xp}; hence, it is a model of non-perturbative mass gap generation (the list of such gapped 2D gauge theories can be found in \cite{Delmastro:2021otj}). 
Surprisingly however, when the adjoint fermion is massless, the fundamental Wilson loop does not obey an area law 
\cite{Gross:1995bp,Gross:1997mx,Komargodski:2020mxz,Dempsey:2021xpf}.
For these and other reasons, the model with one adjoint Majorana fermion has been a nice playground for exploring various interesting phenomena. They include the existence of 
different ``universes,'' i.e.~sectors of the Hilbert space distinguished by the eigenvalue of a one-form symmetry generator \cite{Witten:1978ka,Smilga:1994hc,Lenz:1994du,Cherman:2019hbq,Komargodski:2020mxz}, spontaneous breaking of a discrete chiral symmetry \cite{Smilga:1994hc}, confinement vs.~screening \cite{Gross:1995bp,Gross:1997mx,Cherman:2019hbq,Komargodski:2020mxz,Dempsey:2021xpf}, 
and, more recently, the role of non-invertible symmetries \cite{Komargodski:2020mxz}. 

Following 't Hooft's solution of the large $N_c$ limit of $\grSU(N_c)$ gauge theory with fundamental fermions \cite{tHooft:1974pnl}, 
the bound state spectrum of adjoint QCD$_2$ was studied using light-cone quantization. By now there are quite precise estimates for the masses of the low-lying bound states 
(for recent progress, see \cite{Dempsey:2021xpf,Dempsey:2022uie,Trittmann:2023dar}). However, it is difficult to study the vacuum structure in the light-cone approach.  In addition, it is not clear whether the results obtained from the light-cone approach can fully capture all the universes of the theory.  For these reasons, it is very useful to study adjoint QCD$_2$ on a spatial circle.\footnote{In the small circle limit, where the theory becomes weakly coupled,
the Hamiltonian approach to adjoint QCD$_2$ was implemented in \cite{Lenz:1994du,Cherman:2019hbq}.} In this paper, we introduce a lattice Hamiltonian formulation of this model, which builds both on the Kogut-Susskind approach to lattice gauge theory \cite{Kogut:1974ag} and on the lattice formulation of relativistic Majorana fermions
\cite{Kitaev:2000nmw, Seiberg:2023cdc}. We show that our lattice model
exhibits a number of desirable features, such as the existence of different universes at any lattice spacing. For even $N_c$, we will see that the translation by one lattice site relates different universes in a way that implies the vanishing of the $\frac{N_c}{2}$-string tension.\footnote{The vanishing of the $\frac{N_c}{2}$-string tension was established using anomaly arguments in the continuum theory in \cite{Cherman:2019hbq}.} This is somewhat analogous to the $n$-flavor Schwinger model, where it was recently shown \cite{Dempsey:2022nys, Dempsey:2023gib} that the one-site translation involves a discrete axial transformation in the continuum limit and, when $n$ is odd, changes the theta-angle by $\pi$. 

For $N_c=2$,  there are two distinct universes, with the non-trivial universe corresponding to excitations on top of a chromoelectric flux tube in the fundamental representation. We carry out a numerical study using exact diagonalization and demonstrate good convergence to the continuum limit by comparing the spectrum we obtain with that obtained using light-cone quantization \cite{Dempsey:2022uie}. We also confirm the presence of a non-vanishing fermion bilinear condensate, $\braket{\tr \bar\psi \psi}$, and calculate its numerical value.  Apart from the numerical challenge of an exponential growth of the number of states, the main difficulty we overcome is developing a formulation of our lattice model purely in terms of the gauge-invariant states.\footnote{See also \cite{Hamer:1981yq,Banuls:2017ena} for gauge-invariant formulations of lattice models for $\grSU(2)$ QCD$_2$ with fundamental fermions.}

The rest of this paper is organized as follows.  In Section~\ref{CONTINUUM} we start with a brief review of the $\grSU(N_c)$ adjoint QCD$_2$ theory in the continuum and a few properties of the spectrum as obtained from DLCQ\@.  In Section~\ref{LATTICE} we introduce our lattice model, explain its relation to the continuum, and develop a formulation involving only gauge-invariant states and observables.  In Section~\ref{SYMMETRIES} we discuss the symmetries of the continuum and lattice model and explore some of their consequences.  While so far the discussion is for general $N_c$, in the rest of the paper we focus on the simpler case $N_c = 2$.  In Section~\ref{APPROXIMATIONS}, we use the lattice strong coupling expansion to estimate the energy of the lowest-lying state and obtain good agreement with the DLCQ.   Section~\ref{NUMERICS} contains results for the spectrum and other observables using exact numerical diagonalization.  We end with a discussion of our results in Section~\ref{DISCUSSION}. Technical details are relegated to the appendices.

\section{Continuum Theory}
\label{CONTINUUM}

Let us start by reviewing a few facts about the $\grSU(N_c)$ gauge theory coupled to an adjoint Majorana fermion $\psi$.  The Lagrangian density is
 \es{eq:continuum_Lagrangian}{
	\mathcal{L} = \tr\left(-\frac{1}{2\gym^2} F_{\mu\nu}F^{\mu\nu} 
	 + i\bar\psi \gamma^\mu D_\mu \psi - m\bar\psi \psi\right) \,,
 }
where $F_{\mu\nu} = \partial_\mu A_\nu - \partial_\nu A_\mu -i[A_\mu, A_\nu]$ is the gauge field strength, and the covariant derivative is defined to be $D_\mu \psi = \partial_\mu \psi -i [A_\mu, \psi]$.  For an adjoint-valued field $X$ (such as $A_\mu$, $\psi$, or $F_{\mu\nu}$), we can write $X = X^A T^A$, where the $T^A$ are the Hermitian generators of $\grSU(N_c)$ normalized so that $\tr\left(T^A T^B\right) = \frac{1}{2}\delta^{AB}$, with $A, B = 1, \ldots, N_c^2-1$.  

The definition of the gauge coupling constant used here differs from the convention used in much of the light-cone quantization literature, for instance in \cite{Dalley:1992yy,Kutasov:1993gq,Bhanot:1993xp,Dempsey:2021xpf,Dempsey:2022uie,Trittmann:2023dar}; the gauge coupling used there is $\gDLCQ = \gym / \sqrt{2}$. We will use $\gym$ as defined in \eqref{eq:continuum_Lagrangian} throughout this paper.  Furthermore, let us take $\gamma^0 = \sigma_2$, $\gamma^1 = -i \sigma_3$, and $\gamma^5 = \sigma_1$.  In these conventions, the Majorana spinor $\psi$ is real and consequently $\bar \psi = \psi^T \gamma^0$, with the transpose acting only on the spinor indices.

As pointed out in \cite{Cherman:2019hbq}, it is of further interest to consider a modified model where the double-trace 4-fermion interaction of Gross-Neveu (GN) type is present:
\begin{equation}
\label{fourferm}
\delta \mathcal{L}_{\rm GN} = \kappa (\tr \bar \psi \psi)^2
\ .
\end{equation}
This term, which is forbidden by the super-renormalizibility of the theory (\ref{eq:continuum_Lagrangian}), as well as by a non-invertible symmetry \cite{Komargodski:2020mxz}, can make the model confining even when the adjoint mass vanishes \cite{Cherman:2019hbq}. However, the resulting model has rather different UV properties from the basic model with $\kappa=0$, because the coupling $\kappa$ undergoes a logarithmic running \cite{Cherman:2024on}. To distinguish the model including the 4-fermion term from the basic adjoint QCD$_2$, one may refer to it as ``adjoint GN-QCD$_2$."  Our numerical diagonalizations in the $N_c = 2$ case appear to be consistent with $\kappa = 0$.

The bound state spectrum of the adjoint QCD$_2$ theory with gauge group $\grSU(N_c)$ has been studied using the method of discretized light-cone quantization (DLCQ),
which was introduced in \cite{Pauli:1985ps}.
In the large $N_c$ limit one can make a restriction to the single-trace states, which simplifies the calculations  \cite{Bhanot:1993xp,Demeterfi:1993rs,Boorstein:1993nd,Kutasov:1993gq,Kutasov:1994xq,Gross:1995bp,Dalley:1992yy,Gross:1997mx,Katz:2013qua,Dempsey:2021xpf,Popov:2022vud}.
Some results at finite $N_c$
are also available \cite{Antonuccio:1998uz, Dempsey:2022uie}.

A salient feature of the DLCQ spectrum is that it is gapped for all $N_c$ and $m$.  In particular, it is gapped at $m=0$, with the lightest particle being a fermion of mass $M_f$ and the second lightest being a boson of mass $M_b$. The DLCQ spectrum of the $\grSU(N_c)$ theory was found in \cite{Dempsey:2022uie}:\footnote{For the squared masses of lightest bound states, the coefficients of $\frac{\gym^2 N_c}{2 \pi}$ exhibit weak dependence on~$N_c$.}
 \es{MfMb}{
  m=0: \qquad
   M_f^2 \approx 5.7 \frac{\gym^2 N_c}{2 \pi} \,, \qquad M_b^2 \approx 10.8 \frac{\gym^2 N_c}{2 \pi} \,.
 }
As $m$ increases, $M_f$ grows at a faster rate than $M_b$, and for any $N_c$ the two meet at $m^2 = \frac{\gym^2 N_c}{2\pi}$.  At this value of $m$, the theory becomes supersymmetric \cite{Kutasov:1993gq,Popov:2022vud}. The exact supersymmetry generators of the $(1,1)$ supersymmetry are known in the light-cone quantization of the theory \cite{Boorstein:1993nd,Kutasov:1993gq,Popov:2022vud}.  Note that the light-cone Hamiltonian is invariant under $m \to -m$, so the spectrum obtained from DLCQ will have the same property.  As already mentioned in the Introduction, it is not clear, however, whether the DLCQ spectrum reproduces the spectrum of a single universe of the adjoint QCD$_2$ theory or of all the universes.  We will come back to this question in Section~\ref{NUMERICS}.

\section{Lattice model} 
\label{LATTICE}

We will now formulate a Hamiltonian lattice theory corresponding to a discretization of the continuum theory~\eqref{eq:continuum_Lagrangian}, defined on a spatial circle of length $L$ with periodic boundary conditions both for the fermions and the gauge field.  

\subsection{Lattice Hamiltonian}

Let $N$ be the number of lattice sites, taken to be an even positive integer, and $a = L/N$ be the lattice spacing.  The lattice Majorana fermions live on the lattice sites.  They satisfy the reality condition $\chi_n^{A\dagger} = \chi_n^A$ and the canonical anti-commutation relation
 \es{eq:anti_commutation}{
	\{ \chi_n^A, \chi_m^B\} = \delta_{mn} \delta^{AB} \,.
 }
The lattice analog of the spatial component of the gauge field\footnote{In the Hamiltonian formulation, the time component of the gauge field is eliminated and one has to impose the Gauss law by hand.} are the unitary matrices $U_n$ representing the parallel propagators in the fundamental representation of the gauge group.  The operators $U_n$ live on links where, for each link, the corresponding $U_n$ represents the coordinate of a quantum particle moving on the group manifold.  (For a brief review of a particle moving on a group manifold, see Appendix~\ref{GROUPMANIFOLD}.)  We use the convention where link $n$ joins sites $n$ and $n+1$, with the identification $n \sim n + N$.  The conjugate variables to $U_n$ are the left-acting and right-acting electric fields, which are Lie-algebra valued Hermitian operators with $\grSU(N_c)$ components $L_n^A$ and $R_n^A$, respectively, obeying
 \es{LRCommut}{
	[L_n^A, U_m] = \delta_{nm} T^A U_n \,, \qquad
	 [R_n^A, U_m] = \delta_{nm} U_n T^A \,.
 }
The $L_n^A$ and $R_n^A$ are related via $R_n^A= L_n^B U_n^{BA}$, with $U_n^{AB} \equiv 2 \tr (T^A U_n T^B U_n^{-1})$.

 The lattice Hamiltonian is
 \es{eq:lattice_hamiltonian}{
  H &= \sum_{n=0}^{N-1} \biggl[ 
	\frac{\gym^2 a}{2}  L_n^A L_n^A   
	- \frac{i}{2} \left( a^{-1} + (-1)^n m   \right)  \chi_n^A U_n^{AB} \chi_{n+1}^B  \
	\biggr] \,,
 }
with $n \sim n+N$ identified, as mentioned above.  The Hamiltonian \eqref{eq:lattice_hamiltonian} is invariant under local gauge transformations parameterized by a gauge parameter $V_n$ on every site
 \es{GaugeTransf}{
	 U_n &\to V_n U_n V_{n+1}^{-1} \,, \\
	L_n &\to V_n L_n V_n^{-1} \,, \\ \chi_n &\to V_n \chi_n V_n^{-1} \,.
 }

The Hilbert space that this Hamiltonian acts on is a tensor product of the fermionic Hilbert space and the bosonic one, ${\cal H}_F \otimes {\cal H}_B$.  For the fermionic factor, since we have $N (N_c^2-1)$ Majorana fermions $\chi_n^A$ obeying the Clifford algebra \eqref{eq:anti_commutation}, ${\cal H}_F$ must form a representation of this algebra.  Since $N$ is even, ${\cal H}_F$ will be the direct sum of the two spinor representations of $\mathfrak{so}(N(N_c^2-1))$, for a total dimension of $2^{\frac{N(N_c^2-1)}{2}}$.   The bosonic factor ${\cal H}_B$ is the tensor product of the space of square integrable functions on $\grSU(N_c)$ on each link.  By the Peter-Weyl theorem, on each link we can consider a basis consisting of matrix elements of the $\grSU(N_c)$ group element in all possible irreducible representations.

As in any Hamiltonian lattice gauge theory model, the description of the model is not complete without specifying the Gauss law.  For us, the Gauss law takes the form
 \es{eq:gauss_law}{
	L_n^A - R_{n-1}^A = Q^A_n \,, \qquad \text{for all $n=0,\ldots, N-1$} \,,
 }
where we defined the matter gauge charge $Q_n^A=-\frac i2 f^{ABC} \chi_n^B \chi_n^C$, with $f^{ABC}$ being the structure constants.  The physical states are those states in ${\cal H}_F \otimes {\cal H}_B$ for which \eqref{eq:gauss_law} is obeyed.  Starting in Section~\ref{GAUGEINV}, we will work directly with the gauge-invariant states.

\subsection{Relation to the continuum theory}

Let us now explain how the lattice Hamiltonian introduced in the previous subsection arises from the continuum theory \eqref{eq:continuum_Lagrangian}.  Denoting the lattice sites as $x_n\equiv na$, we make the following staggered identification between the lattice fermion $\chi_n$ and the spinor field $\psi(x)$
\begin{align}\label{eq:fermion_id}
	  \chi_n = \begin{cases}
		\sqrt{2a} \psi_u(x_n) \,, & \text{if $n$ is even} \,, \\
		\sqrt{2a}  \psi_d(x_n) \,, & \text{if $n$ is odd}  \,,
	\end{cases}  \qquad 
	\text{where }  \psi(x) = \begin{pmatrix}
		\psi_u(x) \\
		\psi_d(x)
	\end{pmatrix} \,,
\end{align}
where the upper and lower components of the fermions are discretized on alternating lattice sites.   For the gauge variables, we have link variables $U_n$ and site variables $\phi_n$ that are related, respectively, to the spatial and time components of the gauge field.  
 \es{eq:gauge_id}{
	  U_n &= e^{- i a A_1(x_{n})}  \,, \qquad \phi_n = A_0(x_n) \,.
 }
It is then straightforward to check that the lattice Lagrangian 
 \es{DiscLag}{
  L &= \sum_{n=0}^{N-1} \tr \Biggl[ 
    \frac{1}{g^2 a}  \left(i \dot U_n U_n^{-1} + \phi_n - U_n \phi_{n+1} U_n^{-1} \right)^2   
     +i \chi_n   \dot \chi_n  -  \phi_n [\chi_n, \chi_n] \\
     &{}+ \frac{i}{a} \chi_n U_n \chi_{n+1} U_n^{-1}
      + im (-1)^n \chi_n U_n \chi_{n+1} U_n^{-1}
  \Biggr] 
 }
is an approximation to \eqref{eq:continuum_Lagrangian} as $a \to 0$.

One can introduce the left-acting electric fields $L_n^A = -\frac{2}{g^2 a} \tr \biggl[ T^A \bigl(i \dot U_n U_n^{-1} + \phi_n - U_n \phi_{n+1} U_n^{-1} \bigr) \biggr]$ as the canonically-conjugate variables to the angles parameterizing $U_n$, and then pass to the Hamiltonian
  \es{DiscHamComponents}{
   H &= \sum_{n=0}^{N-1}  \biggl( 
    \frac{g^2 a}{2}  L_n^A L_n^A  
      - \frac{i}{2a} \chi_n^A U_n^{AB} \chi_{n+1}^B 
      - \frac{im}{2} (-1)^n \chi_n^A U_n^{AB}  \chi_{n+1}^B 
  \biggr) \\
  {}&+\sum_{n=0}^{N-1} \phi_n^A \left(L_n^A - R_{n-1}^A - Q_n^A \right)  \,.
 } 
It is then clear that on the gauge-invariant subspace where \eqref{eq:gauss_law} is obeyed, our Hamiltonian reduces to \eqref{eq:lattice_hamiltonian}.

If we want to study the more general space of models which includes the 4-fermion term (\ref{fourferm}), we should generalize this lattice Hamiltonian. The simplest corresponding term on the lattice appears to be 
\begin{equation}
\delta H=- \kappa_{\rm lat} \sum_{n=0}^{N-1} (\tr \chi_n U_n \chi_{n+1} U_n^{-1})^2 \,.
\end{equation}
 Even if we are interested in the basic adjoint QCD$_2$ model without the 4-fermion term in the continuum, it is possible that on the lattice $\kappa_{\rm lat}$ needs to be turned on and appropriately tuned in the large $N$ limit. It is also possible that the sign of $\kappa_{\rm lat}$ induced by the lattice regulator is such that the 4-fermion coupling flows to zero at long distances. The study of these issues is beyond the scope of this paper.

\subsection{General properties of gauge-invariant states}
\label{GAUGEINVSTATES}

As already mentioned, the $2^{N(N_c^2-1)/2}$-dimensional fermionic Hilbert space ${\cal H}_F$ is a sum of the two spinor representations of $\mathfrak{so}(N(N_c^2-1))$.  These two spinor representations are distinguished by the eigenvalues under the $\mathfrak{so}(N(N_c^2-1))$ chirality matrix
 \es{FDef}{
  {\cal F} =  {\cal F}_0 {\cal F}_1 \cdots {\cal F}_{N-1}\,,  \qquad {\cal F}_n \equiv (2i)^{(N_c^2 - 1)/2} \chi_n^1 \chi_n^2 \cdots \chi_n^{N_c^2-1} \,.
 }
Up to an overall sign which is a matter of convention, the non-trivial normalization in \eqref{FDef} ensures that ${\cal F} = {\cal F}^\dagger$ and ${\cal F}^2 = 1$.  The operator ${\cal F}$ should be identified with the fermion parity operator, because under conjugation by it the lattice fermions $\chi_n^A$ change sign:  ${\cal F} \chi_n^A {\cal F}^{-1} = - \chi_n^A$.  ${\cal F}$ can thus be used to split the Hilbert space into bosonic and fermionic states.

The fermionic operators $\chi_n^A$ transform in the vector representation of $\mathfrak{so}(N(N_c^2-1))$.  In order to impose the Gauss law \eqref{eq:gauss_law}, we need to know how the states transform under the charge operators $Q_n^A$, which generate an $\mathfrak{su}(N_c)^N$ subalgebra of $\mathfrak{so}(N(N_c^2-1))$.  The embedding of $\mathfrak{su}(N_c)^N$ into $\mathfrak{so}(N(N_c^2-1))$ is such that the vector representation of $\mathfrak{so}(N(N_c^2-1))$ decomposes as the direct sum
 \es{DirSumchi}{
  (\text{{\bf adj}}, {\bf 1}, {\bf 1}, \ldots) \oplus ({\bf 1}, \text{{\bf adj}}, {\bf 1}, \ldots) \oplus ({\bf 1}, {\bf 1}, \text{{\bf adj}}, \ldots) + \cdots \,,
  }
where $\text{{\bf adj}}$ is the adjoint representation of $\mathfrak{su}(N_c)$.  The decomposition \eqref{DirSumchi} follows from the fact that the $\chi_n^A$ transform in the adjoint of the $\mathfrak{su}(N_c)$ factor for site $n$. 

A group theory exercise (see Appendix \ref{app:rep_decomp}) shows that under $\mathfrak{su}(N_c)^N$, the fermionic Hilbert space ${\cal H}_F$ decomposes as 
 \es{Decomp}{
  {\bf 2^{N(N_c^2-1)/2} } = 2^{N(N_c-1)/2} ( {\bf R}, {\bf R}, \ldots, {\bf R}) \,,
 }
where ${\bf R}$ is the $\mathfrak{su}(N_c)$ representation with Dynkin label $[111\ldots 1]$.  The  Young diagram of ${\bf R}$ consists of one column of each length ranging from $1$ to $N_c-1$, 
 \es{Diagram}{
  {\bf R} = {\tiny
  \begin{ytableau}
 {}& {} & {} & \none[\cdots] & {} & {} & {} \\
 {}& {} & {} & \none[\cdots] & {} & {} \\
 {}& {} & {} & \none[\cdots] & {} \\ 
 \none[\raisebox{-0.03in}{$\vdots$}] & \none[\raisebox{-0.03in}{$\vdots$}] & \none[\raisebox{-0.03in}{$\vdots$}] & \none[\raisebox{-0.03in}{$\iddots$}] \\
 {} & {} & {} \\
 {} & {} \\
 {}
\end{ytableau}}
 \raisebox{0in}{$\left. \rule{0pt}{1.5cm} \right\}\text{$N_c-1$ rows}$} 
\,, \qquad
 \dim {\bf R} = 2^{N_c (N_c-1)/2}  \,,
 }
and it has dimension $2^{N_c (N_c-1)/2}$.  We can interpret \eqref{Decomp} as saying that ${\cal H}_F$ can be realized as a tensor product 
 \es{HFTensorProd}{
  {\cal H}_F = {\cal H}_\text{qubits} \otimes {\cal H}_\text{spins}
 } 
of a vector space ${\cal H}_\text{qubits}$ of $N(N_c-1)/2$ qubits, which has dimension $2^{N(N_c-1)/2}$, and the Hilbert space ${\cal H}_\text{spins}$ of $N$ $\grSU(N_c)$ spins where each site hosts representation ${\bf R}$ of $\mathfrak{su}(N_c)$.  The $N_c$-ality of the representation ${\bf R}$ is
 \es{Nality}{
  \text{$N_c$-ality of ${\bf R}$} = \frac{N_c(N_c-1)}{2} \  (\text{mod $N_c$})
  = \begin{cases}
   \frac{N_c}{2} & \text{if $N_c$ is even} \,, \\
   0 & \text{if $N_c$ is odd} \,,
  \end{cases}
 }
a fact that will be useful later.

For example, when $N_c = 2$, ${\cal H}_\text{qubits}$ is the Hilbert space of $N/2$ qubits, and ${\bf R} = {\bf 2}$ is the spin-$1/2$ representation of $\grSU(2)$.  In this case, we will make the decomposition \eqref{HFTensorProd} more explicit in the next subsection.  When $N_c = 3$, ${\cal H}_\text{qubits}$ is the Hilbert space of $N$ qubits, and ${\bf R} = {\bf 8}$ is the adjoint representation of $\grSU(3)$.  

To construct gauge-invariant states, note that ${\cal H}_\text{qubits}$ is invariant under $\mathfrak{su}(N_c)^N$, so only the ${\cal H}_\text{spins}$ factor participates non-trivially in this construction.  In other words, the gauge-invariant sector of the Hilbert space takes the form
 \es{HGaugeInv}{
  {\cal H} = {\cal H}_\text{qubits} \otimes {\cal H}' \,, \qquad
   {\cal H}' \subset {\cal H}_\text{spins} \otimes {\cal H}_B
 }
where ${\cal H}_B$ is the bosonic Hilbert space.  

The construction of ${\cal H}'$ is as follows.  Let us consider a basis for ${\cal H}_\text{spins}$ to be
 \es{basisHspinsGen}{
  \text{basis for ${\cal H}_\text{spins}$:} \qquad
   \ket{\text{{\bf R}}, \vec{m}_0} \ket{\text{{\bf R}}, \vec{m}_1}   \cdots \ket{\text{{\bf R}}, \vec{m}_{N-1}}
 }
where $\vec{m}_n$ is a multi-index used to label the states of representation $\text{{\bf R}}$  on site $n$.  As already mentioned, on each link $n$ the bosonic Hilbert space ${\cal H}_B$ is that of $L^2$-functions on $\grSU(N_c)$.  Let $\ket{{\bf r}_n, \vec{\mathfrak{m}}_{nL}, \vec{\mathfrak{m}}_{nR}}$ be a basis for this Hilbert space, with ${\bf r}_n$ being an irrep of $\mathfrak{su}(N_c)$ and $\vec{\mathfrak{m}}_{nL}$ and $\vec{\mathfrak{m}}_{nR}$ being multi-indices each labeling the states in this representation.  A basis for the bosonic Hilbert space ${\cal H}_B$ on all $N$ links is then
 \es{bosonicBasis}{
  \text{basis for ${\cal H}_B$:} \qquad \ket{{\bf r}_0, \vec{\mathfrak{m}}_{0L}, \vec{\mathfrak{m}}_{0R} } \ket{{\bf r}_1, \vec{\mathfrak{m}}_{1L}, \vec{\mathfrak{m}}_{1R} }  \cdots \ket{{\bf r}_{N-1}, \vec{\mathfrak{m}}_{(N-1)L}, \vec{\mathfrak{m}}_{(N-1)R} }  \,.
 }
A basis for the gauge invariant subspace ${\cal H}'$ is then of the form
 \es{GaugeInvGen}{
  \sum_{\vec{m}_n, \vec{\mathfrak{m}}_{nL}, \vec{\mathfrak{m}}_{nR}}   \prod_{n=0}^{N-1} \ket{{\bf R}, \vec{m}_n} \otimes \prod_{n=0}^{N-1} \left( C^{{\bf r}_{n-1} {\bf R} {\bf r}_n}_{\vec{\mathfrak{m}}_{(n-1)R} \vec{m}_n \vec{\mathfrak{m}}_{nL}}  \frac{\ket{{\bf r}_n, \vec{\mathfrak{m}}_{nL}, \vec{\mathfrak{m}}_{nR} }}{\dim {\bf r}_n} \right)  \,,
 }
where $C^{{\bf r}_1 {\bf r}_2 {\bf r}_3}_{\vec{m}_1 \vec{m}_2 \vec{m}_3} \equiv \bkt{ {\bf r}_3 \vec{m}_3}{ {\bf r}_1 \vec{m}_1 {\bf r}_2 \vec{m}_2}$ are $\mathfrak{su}(N_c)$ Clebsch-Gordan coefficients.\footnote{For $N_c > 2$, the Clebsch-Gordan coefficients need an additional index to account for the multiplicity of ${\bf r}_n$ in ${\bf r}_{n-1}\otimes {\bf R}$.}    A basis for the full gauge-invariant Hilbert space ${\cal H}$ is obtained by taking the tensor product of the basis \eqref{GaugeInvGen} for ${\cal H}'$ with a basis for ${\cal H}_\text{qubits}$.  

Note that after fixing ${\bf r}_0$, the set of possible representations ${\bf r}_n$ that can appear is restricted by $N_c$-ality.  The ${\bf r}_n$ must obey the property that for any adjacent links the tensor product ${\bf r}_{n-1} \otimes {\bf R}$ must contain the representation ${\bf r}_n$. Thus, the $N_c$-ality of the representations ${\bf r}_n$ must change by \eqref{Nality} when we move from one site to the next. For odd $N_c$, this means we have $N_c$ universes of the Hilbert space where in each universe the $N_c$-ality of all the link irreps is the same.  When $N_c$ is even, we also have $N_c$ universes, but in each universe the $N_c$-ality of the link irreps alternates on even and odd links between values that differ by $N_c /2\pmod{N_c}$.  This interplay between the $N_c$-ality and the translation by one site has interesting consequences that we will explore in more detail in Section~\ref{SYMMETRIES}.

\subsection{Gauge-invariant formulation for $N_c = 2$}
\label{GAUGEINV}

Let us now specialize the general discussion from the previous subsections to the case $N_c = 2$, and further determine the action of the Hamiltonian on the gauge-invariant subspace ${\cal H}'$.  For $N_c = 2$, we can take $T^A  = \sigma^A / 2$, where $\sigma^A$ are the Pauli matrices, and $f^{ABC} = \epsilon^{ABC}$, with $A, B, C = 1, 2, 3$.  

The decomposition ${\cal H}_F = {\cal H}_\text{qubits} \otimes {\cal H}_\text{spins}$ in \eqref{Decomp}--\eqref{HFTensorProd} can be made explicit with an appropriate choice of gamma matrices.  In this case ${\cal H}_\text{qubits}$ is the Hilbert space of $N/2$ qubits and ${\cal H}_\text{spins}$ is the Hilbert space of $N$ spin-$1/2$ particles.  Let $X_k$, $Y_k$, $Z_k$ be the Pauli matrix operators acting on the $k$th qubit, $k = 0, \ldots, \frac{N}{2}-1$, and let $S_n^A$ be the $A$th $\grSU(2)$ generator acting on the $n$th spin as $\sigma^A/2$, with $n = 0, \ldots, N-1$.  With this notation, let us define
 \es{eq:majoranas}{
  \chi_{2k}^A &= \sqrt{2} \left( Z_0 Z_1 \cdots Z_{k-1} X_k\right) \otimes S_{2k}^A  \,, \\
  \chi_{2k+1}^A &= \sqrt{2} \left( Z_0 Z_1 \cdots Z_{k-1} Y_k\right) \otimes S_{2k+1}^A  \,.
 }
One can check that these operators obey the correct anti-commutation relations \eqref{eq:anti_commutation}.  This Clifford algebra representation has the nice property that the $\grSU(2)$ generators on site $n$ $Q^A_n = - \frac{i}{2} \epsilon^{ABC} \chi_n^B \chi_n^C$ and the fermion parity operator ${\cal F}$ defined in \eqref{FDef} each act in only one of the two factors of the tensor product and take the simple forms
 \es{QF}{
    Q^A_n = \mathds{1} \otimes S_n^A \,, \qquad
    {\cal F} = Z_0 Z_1 \cdots Z_{\frac{N}{2} -1} \otimes \mathds{1} \,.
 }
Thus, the $\grSU(2)$ degrees of freedom are carried by the spins, while the fermion parity is carried by the qubits.   In particular, the fact that each spin transforms as a doublet under its corresponding $\grSU(2)$ and the qubits are $\grSU(2)$-invariant confirms the decomposition \eqref{Decomp}.  

The construction of the gauge-invariant states is a simple specialization of the discussion of the previous subsection, with the replacements ${\bf R} \to \half$, ${\bf r} \to \ell$, $\vec{m} \to m$, and $\vec{\mathfrak{m}} \to \mathfrak{m}$, reflecting the fact that $\mathfrak{su}(2)$ representations are labeled by the spin $\ell$ and there is only one quantum number (the magnetic quantum number) labeling the states of these representations.  Thus, we have the bases
 \es{Bases}{
  \text{basis for ${\cal H}_\text{qubits}$}:& \qquad \ket{s_0 s_1 \ldots s_{\frac{N}{2}-1}} \,, \\
  \text{basis for ${\cal H}_\text{spins}$}:& \qquad \ket{\half, m_0} \ket{\half, m_1} \cdots \ket{\half, m_{N-1}} \,, \\
  \text{basis for ${\cal H}_B$}:& \qquad \ket{\ell_0, \mathfrak{m}_{0L}, \mathfrak{m}_{0R} } \ket{\ell_1, \mathfrak{m}_{1L}, \mathfrak{m}_{1R} }  \cdots \ket{\ell_{N-1}, \mathfrak{m}_{(N-1)L}, \mathfrak{m}_{(N-1)R} }
 }
where $s_k \in\{-1, 1\} $ is the eigenvalue of $Z_k$,  $m_n \in \{-\half, \half\} $ is the eigenvalue of $S_n^3$,  $\ell_n=0,\frac{1}{2},1,\ldots$, and both $\mathfrak{m}_{nL}$ and $\mathfrak{m}_{nR}$ range from $-\ell_n$ to $\ell_n$ in integer steps.  (In the position representation, the wavefunction on the group manifold associated with $\ket{\ell_n, \mathfrak{m}_{nL}, \mathfrak{m}_{nR}}$ is $\Psi_{\ell_n, \mathfrak{m}_{nL}, \mathfrak{m}_{nR}}(U_n)$, as defined in \eqref{BasisStates}.)    The gauge-invariant states in ${\cal H}$ are uniquely labeled by the qubit quantum numbers $(s_0, s_1, \ldots, s_{\frac{N}{2}-1})$ for the states in ${\cal H}_\text{qubits}$ and by a string of $\grSU(2)$ angular momenta $(\ell_0, \ell_1, \ldots, \ell_{N-1})$ for the representations on the links for the states in ${\cal H}'$: 
 \es{GaugeInv}{
  &\ket{s_0 \ldots s_{\frac{N}{2}-1}} \otimes \ket{\ell_0 \ldots \ell_{N-1}} \\
  &\text{with } \ket{\ell_0 \ldots \ell_{N-1}} = \sum_{m_n, \mathfrak{m}_{nL}, \mathfrak{m}_{nR}}   \prod_{n=0}^{N-1} \ket{\half, m_n} \otimes \prod_{n=0}^{N-1} \left( C^{\ell_{n-1} \frac{1}{2} \ell_n}_{\mathfrak{m}_{(n-1)R} m_n \mathfrak{m}_{nL}}  \frac{\ket{\ell_n, \mathfrak{m}_{nL}, \mathfrak{m}_{nR} }}{\sqrt{2 \ell_n + 1}} \right)  \,,
 }
with the condition that
 \es{eq:link_condition}{
	|\ell_{n+1} - \ell_n| = \frac{1}{2} \,, \qquad\text{and}\qquad \ell_{-1} = \ell_{N-1} \,.
 }

The space of gauge configurations is of course infinite, but we can truncate it by requiring $\ell_n \leq \lmax$ for all $n$, for some $\lmax$.  In Table~\ref{tab:state_counts}, we give the sizes of the truncated Hilbert space for different $N$ and $\lmax$.
\begin{table}
	\begin{tabularx}{\linewidth}{p{1.7cm}*{5}{R}}
		\toprule
		$\lmax\backslash N$ & 4 & 6 & 8 & 10 & 12 \\
		\midrule
		2 & 40 & 224 & 1312 & 7808 & 46720 \\
		3 & 64 & 384 & 2432 & 15872 & 105472 \\
		4 & 88 & 544 & 3552 & 23936 & 164608 \\
		\bottomrule
	\end{tabularx}
	\caption{The dimensions of the Hilbert space for various values of $N$ and $\lmax$. For the plots in this paper we use up to $N = 12$ and $\lmax = 4$.}
	\label{tab:state_counts}
\end{table}

Having established a basis for the gauge invariant subspace, we need to determine the action of the Hamiltonian~\eqref{eq:lattice_hamiltonian} on these states.  The gauge kinetic term $H_\text{gauge} = \frac{\gym^2 a}{2} L_n^A L_n^A$ is diagonal in this basis, and it acts only on the ${\cal H'}$ factor:
\begin{equation}\label{eq:gauge_kin_action}
	H_\text{gauge} \ket{\ell_0 \ldots \ell_{N-1}} = \left(\frac{\gym^2 a}{2} \sum_{n=0}^{N-1} \ell_n (\ell_n + 1)\right)\ket{\ell_0 \ldots \ell_{N-1}} \,.
\end{equation}
For each link, the second term in \eqref{eq:lattice_hamiltonian} is proportional to the operator $-\frac{i}{2} \chi_n^A U_n^{AB} \chi_{n+1}^B$.  This operator acts on both factors of the Hilbert space non-trivially.  Based on whether $n$ is even or odd, the action is:
\begin{subequations}\label{eq:pauli_hopping}
	\begin{align}
		-\frac{i}{2}\chi^A_{2k} U_{2k}^{AB} \chi^B_{2k+1} &= Z_k \otimes {\cal O}_{2k}  \,, \qquad
		 \qquad \qquad {\cal O}_n \equiv S_n^A U_n^{AB} S_{n+1}^B \,, \label{eq:even_pauli}\\
		-\frac{i}{2}\chi^A_{2k+1} U_{2k+1}^{AB} \chi^B_{2k+2} &= (-{\cal F})^{\delta_{k, \frac{N}{2}-1}} X_k X_{k+1} \otimes {\cal O}_{2k+1}  \,. \label{eq:odd_pauli}
	\end{align}
\end{subequations}
It is straightforward to determine the actions of the first factors in \eqref{eq:pauli_hopping} on the qubits since the basis states $\ket{s_0 \ldots s_{\frac{N}{2} - 1}}$ are eigenstates of $Z_k$ with eigenvalue $s_k$, and $X_k$ simply flips the sign of $s_k$.  The action of ${\cal O}_n$ on the states of ${\cal H}'$ is harder to determine, but we show in Appendix~\ref{GAUGEINVAPPENDIX} that this action is
 \es{OnAction}{
  {\cal O}_n \ket{\ell_0 \ldots \ell_{N-1}} = \sum_{\ell_n' \in \{\ell_n-1, \ell_n, \ell_n+1\}}
   f(\ell_{n-1}, \ell_{n+1};  \ell_n', \ell_n) \ket{\ell_0 \ldots \ell_{n-1} \ell_n' \ell_{n+1} \ldots \ell_{N-1}} \,,
 }
with the expression for $f$ given in~\eqref{Gotf}.

It would seem that this gauge invariant formulation of the lattice theory does not share the symmetry of translating by two sites that is manifest in the Hamiltonian~\eqref{eq:lattice_hamiltonian}. However, the insertion of the operator $(-{\cal F})$ in $\chi_{N-1}^A U_{N-1}^{AB} \chi^B_0$ is unitarily equivalent to an insertion of $(-{\cal F})$ in any other hopping term $\chi^A_{2k-1}U_{2k} \chi^B_{2k}$. Specifically, the unitary that moves $(-{\cal F})$ between a hopping term with a given $k$ and that with $k+1$ is implemented by 
\begin{align}
		\mathcal{U}_{k} = \frac{1}{2}(\mathds{1}-{\cal F})+\frac{1}{2}Z_k({\cal F}+\mathds{1}) \,.
\end{align}
If we denote the naive translation by two sites as $T_{2}'$ then the genuine symmetry of the Hamiltonian $T_2$ can be written as $T_2 = \mathcal{U}_{N/2} T_2'$.

\section{Symmetries}
\label{SYMMETRIES}

In this section, we discuss the symmetries of the adjoint QCD$_2$ theory.  We start with a discussion of the continuum theory \eqref{eq:continuum_Lagrangian} in Section~\ref{SYMCONT}, and then in Section~\ref{SYMLATTICE} we proceed with an analogous discussion for the lattice model \eqref{eq:lattice_hamiltonian} introduced in the previous section.   As we will see, the symmetries of the two models mirror each other very closely.  The consequences of the various anomalies on the spectrum, which we discuss in Section~\ref{CONSEQUENCES} below, will be the same in the two cases.

\subsection{Symmetries of the continuum theory}
\label{SYMCONT}

As shown in \cite{Cherman:2019hbq}, the internal (non-space-time) symmetries of the continuum theory \eqref{eq:continuum_Lagrangian} are
 \es{SymmsCont}{
  m \neq 0: \qquad \begin{cases}
   \Z_2^{[1]} \times (\Z_2)_F   \,, & \text{for $N_c=2$} \,, \\
   \left[ \Z_{N_c}^{[1]} \rtimes (\Z_2)_C \right] \times (\Z_2)_F  \,, & \text{for $N_c>2$}  \,,
  \end{cases}
 }
for generic $m \neq 0$, and
 \es{SymmsContMassless}{
  m = 0: \qquad \begin{cases}
   \Z_2^{[1]} \times (\Z_2)_F \times (\Z_2)_\chi  \,, & \text{for $N_c=2$} \,, \\
   \left[ \Z_{N_c}^{[1]} \rtimes (\Z_2)_C \right] \times (\Z_2)_F \times (\Z_2)_\chi  \,, & \text{for $N_c>2$}  \,,
  \end{cases}
 }
for $m=0$.  In particular, for every $N_c$ we have a fermion parity symmetry $(\Z_2)_F$ and a discrete axial symmetry $(\Z_2)_\chi$ (present only when $m=0$) that act by sending $\psi \to -\psi$ and $\psi \to \gamma_5 \psi$, respectively, while leaving $A_\mu$ invariant.  We also have a $\Z_{N_c}$ one-form symmetry $\Z_{N_c}^{[1]}$ corresponding to the center symmetry of $\grSU(N_c)$, under which the fundamental Wilson lines have charge $e^{2 \pi i / N_c}$.  Lastly, for $N_c > 2$, we also have a charge conjugation symmetry $(\Z_2)_C$, which acts by sending $\psi \to \psi^T$ and $A_\mu \to -A_\mu^T$, where the transpose acts on the generators in the fundamental representation.  (We will write this transformation more explicitly later in the lattice model.)  This charge conjugation symmetry does not commute with the one-form center symmetry because it takes a fundamental Wilson line to an anti-fundamental one, so the symmetry group involves a semi-direct product between these two symmetries.  Lastly, when $N_c = 2$, the charge conjugation symmetry is absent because $\psi \to \psi^T$ and $A_\mu \to -A_\mu^T$ is a gauge transformation.

Let us restrict ourselves to the theory compactified on a spatial circle with periodic boundary conditions, because this is what the lattice model introduced in the previous section approximates.\footnote{It is possible to also study both the continuum theory and the lattice model with anti-periodic boundary conditions for the fermions.}   Let us denote the generators of $\Z_{N_c}^{[1]}$, $(\Z_2)_C$, $(\Z_2)_F$, and $(\Z_2)_\chi$ by $\widehat{\cal U}(x)$, $\widehat{\cal C}$, $\widehat{\cal F}$, and $\widehat{\cal V}$, respectively, where $x$ is the coordinate parameterizing the circle. (We use hats when denoting the unitary operators in the continuum theory in order to distinguish these operators from those in the lattice model, for which we will not use hats.) These are unitary operators that act on the Hilbert space, and they can be discussed regardless of whether the corresponding transformations they implement are symmetries of the theory or not.  As shown in \cite{Cherman:2019hbq}, the algebra obeyed by the non-chiral symmetries $\Z_{N_c}^{[1]}$, $(\Z_2)_C$, and $(\Z_2)_F$ is the same at the classical and quantum levels, namely:
 \es{AlgVector}{
  \widehat{{\cal U}}(x)^{N_c} &= 1 \,, \qquad \widehat{{\cal C}}^2 = \widehat{{\cal F}}^2 = 1 \,, \\
  \widehat{{\cal U}}(x) \widehat{{\cal C}} &= \widehat{{\cal C}} \widehat{{\cal U}}(x)^{-1} \,, \qquad
   \widehat{{\cal U}}(x) \widehat{{\cal F}} = \widehat{{\cal F}} \widehat{{\cal U}}(x) \,, \qquad \widehat{{\cal F}} \widehat{{\cal C}} = \widehat{{\cal C}} \widehat{{\cal F}} \,.
 } 
However, the algebra involving the axial symmetry $\widehat{\cal V}$ is realized projectively, and  \cite{Cherman:2019hbq} found that
 \es{AlgAxial}{
  \widehat{\cal V}^2 &= 1\,, \qquad \qquad \qquad 
    \widehat{\cal U}(x) \widehat{\cal V} = (-1)^{N_c -1}  \widehat{\cal V} \widehat{\cal U}(x) \,, \\
 \widehat{\cal F} \widehat{\cal V} &= (-1)^{N_c -1}  \widehat{\cal V} \widehat{\cal F}   \,, \qquad \ \ 
   \widehat{\cal C} \widehat{\cal V} = (-1)^{\frac{(N_c-2)(N_c-1)}{2}}  \widehat{\cal V} \widehat{\cal C}   \,.
 } 
The non-trivial signs in these expressions would not have been present classically and are signals of quantum anomalies.  They show that the algebra of the unitary operators introduced above is realized projectively on the Hilbert space.

Lastly, let us discuss how these unitary operators act on the Hamiltonian.  Let us denote the continuum analog of the Hamiltonian \eqref{eq:lattice_hamiltonian} by $H_m$ so we can keep track of the mass $m$.  Since $\Z_{N_c}^{[1]}$, $(\Z_2)_C$, and $(\Z_2)_F$ are symmetries for all $m$, the corresponding generators commute with the Hamiltonian, or, equivalently, the Hamiltonian is invariant under conjugation by these unitary operators:
 \es{VectHam}{
  \widehat{\cal U}(x) H_m \widehat{\cal U}(x)^{-1} = H_m \,, \qquad
   \widehat{\cal C} H_m \widehat{\cal C}^{-1} = H_m \,, \qquad 
   \widehat{\cal F} H_m \widehat{\cal F}^{-1} = H_m \,.
 }
On the other hand, the $(\Z_2)_\chi$ axial transformation is a symmetry only for $m=0$, so conjugation by it does not leave the Hamiltonian invariant for $m\neq 0$.  Since the operator $\tr \bar \psi \psi$ changes sign under the axial transformation, the conjugation of the Hamiltonian by $\widehat{\cal V}$ has the simple effect of changing the sign of $m$:
 \es{VHam}{
   \widehat{\cal V} H_m \widehat{\cal V}^{-1} = H_{-m} \,.
 }

The relations \eqref{AlgVector}--\eqref{VHam} have important consequences for the spectrum and other observables. (See also the discussion in \cite{Cherman:2019hbq}.)  Let us postpone the discussion of these consequences until after we present the analogous relations to \eqref{AlgVector}--\eqref{VHam} on the lattice, because the consequences will be the same in both cases.

\subsection{Symmetries of the lattice model}
\label{SYMLATTICE}

The lattice model \eqref{eq:lattice_hamiltonian} exhibits a very similar set of symmetries and anomalies as the continuum model.  In particular, there exist unitary operators representing a lattice one-form symmetry ${\cal U}_n$, fermion parity ${\cal F}$, and charge conjugation ${\cal C}$.  The lattice analog of the $(\Z_2)_\chi$ generator will be the translation operator by one lattice site, ${\cal V}$, which in the continuum limit reduces to the product between a $(\Z_2)_\chi$ transformation and an infinitesimal translation.   Let us discuss these unitary operators one by one.

\subsubsection{Definitions of unitary operators}

The lattice one-form symmetry is defined as follows.   Let ${\cal Z}_n$ be the generator of the $\Z_{N_c}$ center of $\mathfrak{su}(N_c)$ acting on the bosonic Hilbert space on link $n$ (before imposing the Gauss law).  On a basis state $\ket{{\bf r}_n, \vec{\mathfrak{m}}_{nL}, \vec{\mathfrak{m}}_{nR}}$, ${\cal Z}_n$ acts as ${\cal Z}_n \ket{{\bf r}_n, \vec{\mathfrak{m}}_{nL}, \vec{\mathfrak{m}}_{nR}} = e^{2 \pi i (\text{$N_c$-ality of ${\bf r}_n$}) / N_c}\ket{{\bf r}_n, \vec{\mathfrak{m}}_{nL}, \vec{\mathfrak{m}}_{nR}}$.  Clearly, ${\cal Z}_n^{N_c} = 1$ because this relation holds on all basis states.  In term of ${\cal Z}_n$, the generator of the lattice $\Z_{N_c}$ one-form symmetry can be written as
 \es{calULattice}{
  {\cal U}_n = (-1)^{(n+1) (N_c - 1)} {\cal Z}_n \,.
 }
Both ${\cal Z}_n$ and ${\cal U}_n$ commute with the lattice Hamlitonian \eqref{eq:lattice_hamiltonian} because acting with the Hamiltonian does not change the $N_c$-ality of the states.  However, only ${\cal U}_n$ is a one-form symmetry for all $N_c$ because it further obeys the property that it is topological, namely that ${\cal U}_n = {\cal U}_{n+1}$ when acting on gauge-invariant states.  Indeed, this property holds because the $(-1)^{(n+1) (N_c - 1)}$ factor in \eqref{calULattice} accounts for the change in $N_c$-ality when going from one link to the next, as given by combining the Gauss law \eqref{eq:gauss_law} with the $N_c$-ality \eqref{Nality} of the states.   ${\cal U}_n$ generates a $\Z_{N_c}$ symmetry because ${\cal U}_n^{N_c} = (-1)^{(n+1) N_c (N_c - 1)} {\cal Z}_n^{N_c} =  1$ after using ${\cal Z}_n^{N_c} =  1$ and the fact that $N_c(N_c-1)$ is always an even integer.

The fermion parity operator ${\cal F}$ was already defined in \eqref{FDef}.  It commutes with the Hamiltonian and it obeys ${\cal F}^2 = 1$.

The definition of the charge conjugation symmetry operator that most closely resembles the continuum analog is that for which 
 \es{calCAction}{
  {\cal C} \chi_n {\cal C}^{-1} = -\chi_n^T \,, \qquad
   {\cal C} U_n {\cal C}^{-1} = (U^{-1})^T = U^* \,,
 }
where the transpose acts on the color indices when $\chi_n$ and $U_n$ are represented as $N_c \times N_c$ matrices acting in the fundamental representation of $\mathfrak{su}(N_c)$.  The minus sign in the first equation is a matter of convention and can be removed by replacing ${\cal C} \to {\cal F} {\cal C}$. To make \eqref{calCAction} more explicit, let us order the $\mathfrak{su}(N_c)$ generators $T^A$ such that the first ${\cal N}_I = \frac{N_c (N_c-1)}{2}$ generators are represented by pure imaginary anti-symmetric matrices in the fundamental representation, while the last ${\cal N}_R = \frac{(N_c+2) (N_c-1)}{2}$ generators are represented by real traceless symmetric matrices.  In components, the transformations \eqref{calCAction}, as well as the corresponding transformations of the electric fields can then be written as 
 \es{calCAction2}{
  {\cal C} \chi_n^A {\cal C}^{-1} &= -(-1)^{\theta_A} \chi_n^A \,, \qquad 
    {\cal C} U_n^{AB} {\cal C}^{-1} = (-1)^{\theta_A + \theta_B} U_n^{AB} \,, \\
  {\cal C} L_n^{A} {\cal C}^{-1}  &= -(-1)^{\theta_A} L_n^A  \,, \qquad
   {\cal C} R_n^{A} {\cal C}^{-1}  = -(-1)^{\theta_A} R_n^A \,,
  }
 where $\theta_A = 1$ for the anti-symmetric generators and $\theta_A = 0$ for the symmetric ones:
  \es{thetaDef}{
   \theta_A \equiv \begin{cases}
      1 & \text{for $1 \leq A \leq {\cal N}_I$} \,, \\
      0 & \text{for ${\cal N}_I < A \leq {\cal N}_I + {\cal N}_R$} \,.
     \end{cases}
  } 
 When checking that the transformations \eqref{calCAction2} are consistent with the various commutation relations (for instance \eqref{LRUCommut}), it is useful to note that the structure constants $f^{ABC}$ obey the property $(-1)^{\theta_A + \theta_B + \theta_C} f^{ABC} = - f^{ABC}$.  This property can also be used to check that ${\cal C} Q_n^A {\cal C}^{-1} = - (-1)^{\theta_A} Q_n^A$, which implies that the transformation rules \eqref{calCAction2} are also consistent with the Gauss law \eqref{eq:gauss_law}.   Furthermore, one can also see that \eqref{calCAction2} imply that conjugation by ${\cal C}$ leaves the Hamiltonian invariant.  
   
The unitary operator ${\cal C}$ that implements \eqref{calCAction2} can be written as a product of two unitaries ${\cal C} = {\cal C}_F {\cal C}_B$, where ${\cal C}_F$ acts on the fermions and ${\cal C}_B$ on the bosons.  We will not need an expression for ${\cal C}_B$ because for the bosons there are no non-trivial signs that can be generated at the quantum level.  However, let us write ${\cal C}_F$ explicitly because its expression will be needed for computing the algebra of operators:
 \es{CFExplicit}{
  {\cal C}_F = {\cal C}_{F0} {\cal C}_{F1} \cdots {\cal C}_{F, N-1} \,, \qquad
   {\cal C}_{Fn} \equiv (2i)^{\frac{{\cal N}_R}{2}} \chi_n^{{\cal N}_I + 1} \chi_n^{{\cal N}_I + 2} \cdots \chi_n^{{\cal N}_I + {\cal N}_R} \,. 
 }
This definition ensures that ${\cal C}_F^2 = 1$ and ${\cal C}_F = {\cal C}_F^\dagger$, so if ${\cal C}_B$ is defined to obey the same properties, then so will ${\cal C}$.  We will assume this is the case.

Lastly, we also consider the unitary operator ${\cal V}$ that implements lattice translation by one site.  For our purposes it will be enough to know that
 \es{calVExplicit}{
  {\cal V} \chi_n {\cal V}^{-1} = \chi_{n+1} \,, \qquad
   {\cal V} U_n {\cal V}^{-1} = U_{n+1} \,, \qquad
    {\cal V} L_n {\cal V}^{-1} = L_{n+1} \,, \qquad
      {\cal V} R_n {\cal V}^{-1} = R_{n+1} \,,
 }
of course with understanding that the indices obey the identification $n \sim n+N$.  The operator ${\cal V}$ commutes with the Hamiltonian when $m=0$, but when $m \neq 0$ it flips the sign of $m$, just like in the continuum.  Thus, if we denote the lattice Hamiltonian \eqref{eq:lattice_hamiltonian} by $H_m$ in order to keep track of $m$, then ${\cal V} H_m {\cal V}^{-1} = H_{-m}$.  With an appropriate normalization, one can realize ${\cal V}$ on the Hilbert space in a way that ${\cal V}^N = 1$.  This is the main difference between the translation by one site and the $(\Z_2)_\chi$ transformation in the continuum:  while in the continuum $\widehat{\cal V}^2 = 1$, on the lattice we have ${\cal V}^N = 1$ instead. 
In the language of recent papers \cite{Cheng:2022sgb,Seiberg:2023cdc}, the $(\Z_2)_\chi$ symmetry of the continuum theory ``emanates" from the lattice translation by one site, ${\cal V}$.

\subsubsection{Operator algebra}

With the definitions above, we can determine the algebra obeyed by the various unitary operators.  The operators ${\cal U}_n$, ${\cal C}$, and ${\cal F}$ obey the same algebra that their hatted counterparts obey in the continuum (see \eqref{AlgVector}):
 \es{AlgVectorLattice}{
  {\cal U}_n^{N_c} &= 1 \,, \qquad {\cal C}^2 = {\cal F}^2 = 1 \,, \\
  {\cal U}_n {\cal C} &= {\cal C} {\cal U}_n^{-1} \,, \qquad {\cal U}_n {\cal F} = {\cal F} {\cal U}_n \,,
   \qquad {\cal F} {\cal C} = {\cal C} {\cal F} \,.
 } 
Now also including the ${\cal V}$ operator, we have a lattice analog of \eqref{AlgAxial},
 \es{AlgAxialLattice}{
  {\cal V}^N &= 1 \,, \qquad {\cal U}_n {\cal V} = (-1)^{N_c - 1} {\cal V} {\cal U}_n \,, \\
  {\cal F} {\cal V} &= (-1)^{N_c - 1} {\cal V} {\cal F} \,, \qquad
   {\cal C} {\cal V} = (-1)^{\frac{(N_c-2)(N_c-1)}{2}} {\cal V} {\cal C} \,,
 }
with the only difference being that ${\cal V}^N = 1$ instead of $\widehat{\cal V}^2 = 1$, as mentioned above.

Conjugating the Hamiltonian $H_m$ by the four unitary operators, we obtain an exact analog of the continuum relations \eqref{VectHam}--\eqref{VHam}:
 \es{UnitHamLattice}{
  {\cal U}_n H_m {\cal U}_n^{-1} = H_m \,, \qquad {\cal C} H_m {\cal C}^{-1} &= H_m \,, \qquad
   {\cal F} H_m {\cal F}^{-1} = H_m \,, \\
  {\cal V} H_m {\cal V}^{-1} &= H_{-m} \,. 
 }

\subsection{Consequences for the spectrum}
\label{CONSEQUENCES}

The consequences of the relations \eqref{AlgVectorLattice}--\eqref{UnitHamLattice} are precisely the same as those of the continuum relations \eqref{AlgVector}--\eqref{VHam}.  Let us describe these consequences in the language of the lattice model, but the exact same conclusions hold for the spectrum of the continuum theory compactified on a circle with periodic boundary conditions for the fermions.

First, the lattice one-form symmetry can be used to split the Hilbert space into $N_c$ distinct universes.  Let the $p$th universe be the sector of the Hilbert space where ${\cal U}_n = e^{2 \pi i p / N_c}$  ($p$ is an integer identified modulo $N_c$).  Similarly, the fermion parity operator can be used to split the Hilbert space into bosonic and fermionic states based on whether ${\cal F} = +1$ or $-1$.  Since ${\cal U}_n$, ${\cal F}$, and $H_m$ commute, they are simultaneously diagonalizable, so one can consider a basis of eigenstates of $H_m$ that are also eigenstates of ${\cal U}_n$ and ${\cal F}$.  

If we act with ${\cal V}$ on a simultaneous eigenstate $\ket{\psi}$  of ${\cal U}_n$, ${\cal F}$, and $H_m$ with eigenvalues $e^{2 \pi i p / N_c}$, $f$, and $E$, the relations \eqref{AlgAxialLattice}--\eqref{UnitHamLattice} imply that ${\cal V} \ket{\psi}$ is a simultaneous eigenstate of ${\cal U}_n$, ${\cal F}$, and $H_{-m}$ with eigenvalues  $(-1)^{N_c - 1} e^{2 \pi i p / N_c}$, $(-1)^{N_c - 1} f$, and $E$.  This means:
 \begin{itemize}
  \item If $N_c$ is even, then the bosonic/fermionic eigenstates of $H_m$ in the $p$th universe ($0 \leq p < \frac{N_c}{2}$) are exactly degenerate with the fermionic/bosonic eigenstates of $H_{-m}$ in the $(\frac{N_c}{2} + p)$th universe.  Note that this implies an exact degeneracy in the spectrum at $m=0$ between the $p$th and $(\frac{N_c}{2} + p)$th universes.
    \item If $N_c$ is odd, then the bosonic/fermionic eigenstates of $H_m$ in the $p$th universe ($0 \leq p < N_c$) are exactly degenerate with the bosonic/fermionic eigenstates of $H_{-m}$ in the same universe.  Thus, the energy spectrum of each universe is invariant under $m \to -m$.  However, note that this statement does not necessarily imply a degeneracy in the spectrum at $m=0$.
 \end{itemize}
 Also note that since the mass operator ${\cal O}_\text{mass} = \frac{i}{2} \sum_n (-1)^n \chi_n^A U_n^{AB} \chi_{n+1}^B$ changes sign when conjugated by ${\cal V}$, ${\cal V} {\cal O}_\text{mass} {\cal V}^{-1} = - {\cal O}_\text{mass}$, it follows that 
  \es{ExpValSign}{
   \langle \psi | {\cal O}_\text{mass} | \psi \rangle = - \langle {\cal V} \psi | {\cal O}_\text{mass} | {\cal V} \psi \rangle \,.
  }
The same expression holds in the continuum where ${\cal O}_\text{mass} = \tr \bar \psi \psi$.
 
If on the same $\ket{\psi}$ as above we act with ${\cal C}$, the relations \eqref{AlgVectorLattice} and \eqref{UnitHamLattice} imply that ${\cal C} \ket{\psi}$ is a simultaneous eigenstate of ${\cal U}_n$, ${\cal F}$, and $H_{m}$ with eigenvalues  $ e^{-2 \pi i p / N_c}$, $f$, and $E$.  Thus:
 \begin{itemize}
  \item For any $N_c > 2$, the bosonic/fermionic eigenstates of $H_m$ in the $p$th universe are exactly degenerate with the bosonic/fermionic eigenstates of $H_m$ in the $(N_c - p)$th universe.
 \end{itemize}

For $m=0$ and $N_c$ odd, ${\cal V}$ commutes with ${\cal U}_n$, ${\cal F}$, and $H_0$, so the four operators are now simultaneously diagonalizable.  Let $\ket{\psi}$ now be a simultaneous eigenstate of these four operators with eigenvalues $v$, $e^{2 \pi i p / N_c}$, $f$, and $E$.  Then ${\cal C} \ket{\psi}$ is also a simultaneous eigenstate of the four operators, with eigenvalues $(-1)^{\frac{(N_c-2)(N_c-1)}{2}} v$, $e^{-2 \pi i p / N_c}$, $f$, and $E$.  When $N_c = 4k + 1$ for some integer $k$, then $(-1)^{\frac{(N_c - 1)(N_c-2)}{2}} = 1$, and no additional conclusions can be drawn.  When $N_c$ is of the form $N_c = 4k + 3$ for some integer $k$, then $(-1)^{\frac{(N_c - 1)(N_c-2)}{2}} = -1$, and we conclude that the eigenvalue of ${\cal V}$ changes sign upon acting with ${\cal C}$.  Then:
 \begin{itemize}
  \item When $m=0$ and $N_c = 4k+3$ for some integer $k$, the bosonic/fermionic eigenstates of $H_0$ in the $p$th universe ($0\leq p < \frac{N_c}{2}$) with ${\cal V}$-eigenvalue $v$ are exactly degenerate with bosonic/fermionic eigenstates of $H_0$ in the $(N_c - p)$th universe with ${\cal V}$-eigenvalue $-v$.  
  
 \end{itemize}
 
 Lastly, one can make an additional statement about the $p=0$ universe when $N_c = 4k+3$ and any $m$.  When restricted to this universe, ${\cal C}$ acts as a symmetry, so one can simultaneously diagonalize ${\cal F}$, ${\cal C}$, and $H_m$.  Let $f$, $c$, and $E$ be the corresponding eigenvalues of a state $\ket{\psi}$.  Then ${\cal V} \ket{\psi}$ is also in the $p=0$ universe and it is a simultaneous eigenstate of ${\cal F}$, ${\cal C}$, and $H_m$ with eigenvalues $f$, $-c$, and $E$.  Thus:
  \begin{itemize}
   \item When $p=0$ and $N_c = 4k+3$, bosonic/fermionic eigenstates of $H_m$ with $C = \pm 1$ are exactly degenerate with bosonic/fermionic eigenstates of $H_{-m}$ with $C = \mp 1$.  Note this implies that when we further set $m=0$, the spectrum of the $p=0$ universe has an exact double degeneracy between states with $C=+1$ and $C=-1$.
  \end{itemize} 
Note that the exactly degenerate states related by the action of ${\cal V}$ have opposite expectation values for ${\cal O}_\text{mass}$ according to \eqref{ExpValSign}.

\subsection{Supersymmetry}
\label{Supersymmetry}

In addition to the massless point, which is distinguished by the presence of the axial symmetry discussed in the previous sections, adjoint QCD$_2$ has a special point 
\begin{equation}
	m^2 = m^2_\text{SUSY} \equiv \frac{\gym^2 N_c}{2\pi}\ .
\end{equation}
Using the light-cone quantization, it was shown that at this point the model 
exhibits $(1,1)$ supersymmetry \cite{Kutasov:1993gq,Popov:2022vud}. This very interesting result has been checked using DLCQ \cite{Bhanot:1993xp,Boorstein:1993nd,Antonuccio:1998zp,Dempsey:2022uie}. In this section, we briefly discuss the consequences of the supersymmetry for the spectrum of the theory on a (discretized) spatial circle.

When $m = 0$, adjoint QCD$_2$ has several degenerate vacua, which we expect to be split as we turn on the mass. At $m = m_\text{SUSY}$, we expect that the lowest vacuum will be annihilated by the supersymmetry generator, while the other vacua exhibit spontaneous breaking of supersymmetry. Thus, there will be massless Goldstinos in these higher vacua which contain wound flux tubes \cite{Dubovsky:2018dlk}, while the zero energy vacuum preserves supersymmetry.

There are two regimes in which we can understand the ordering of the vacua. For $m\gg \gym$ we can integrate out the adjoint fermion, and so the vacuum energy will be given by the energy of the flux tube wrapping the spatial circle.  This energy is proportional to the lowest value of the quadratic Casimir of a representation with $N_c$-ality $p$.  Thus, the vacua of the universe with $p = 0$ will be the lowest in this limit, followed by $p=1,N_c -1$, followed by $p=2, N_c-2$, etc.  In the opposite limit, when $\abs{m}\ll \gym$, the vacuum energies will be given approximately by $m\langle \mathcal{O}_\text{mass}\rangle$. Thus, at $m = m_\text{SUSY} \sim \gym$, we expect that the vacuum of the $p = 0$ universe that is continuously connected to the vacuum with the most negative VEV of the mass operator at $m = 0$.

For $N_c = 2$, this manifests in a simple manner: there are two universes, and it is the vacuum of the $p = 0$ universe that preserves supersymmetry at $m = \frac{\gym}{\sqrt{\pi}}$. The $p = 1$ universe has a massless Goldstino at this point. This implies also that the $p = 0$ universe has a massless excitation at $m = -\frac{\gym}{\sqrt{\pi}}$. We can see this explicitly with our lattice model for $N_c = 2$, as in Figure \ref{fig:fermion_mass}.
For a slightly more complicated example, we can take $N_c = 3$.  At $m/\gym = \sqrt{\frac{3}{2\pi}}$, we expect to see a massless Goldstino in the $p = 1,2$ vacua (they correspond to having a confining flux tube wound around the circle in one or the other direction), but not in the trivial sector $p=0$.

\section{Strong coupling expansion for $\grSU(2)$}
\label{APPROXIMATIONS}

With the lattice formulation \eqref{eq:lattice_hamiltonian} of adjoint QCD$_2$, we can develop a strong coupling expansion analogous to the one performed extensively for the Schwinger model, for instance in \cite{Banks:1975gq,Hamer:1997dx,Hamer:1982mx}.   The lattice strong coupling expansion is an expansion in $x \equiv 1/ (\gym a)^{2}$.  To approach the continuum limit, one has to extrapolate this expansion to $x \to \infty$, which we will do using appropriate Pad\'e  approximants. 

In this section we set $m=0$\footnote{In the strong coupling expansion for the Schwinger model \cite{Banks:1975gq}, it is possible to include a non-zero fermion mass in the unperturbed Hamiltonian.  This cannot be done for this model because, unlike in the Schwinger model, the fermion mass term does not commute with the gauge kinetic term.} and we study the $p = 0$ universe, which has half-integer $\ell_n$ on even links and integer $\ell_n$ on odd links. (Recall that the two universes are exactly degenerate at $m=0$, so it suffices to focus on one of them.) To facilitate the strong coupling expansion, we rescale the Hamiltonian and then write it as the sum of a diagonal term $h_0$ and a small perturbation $V$:
\begin{align}\label{eq:strong_setup}
	H = \frac{\gym}{\sqrt{x}} h \,, \qquad
	 h = h_0 + x V  \,,
\end{align}
where 
 \es{h0VDefs}{
  h_0 =  \frac{1}{2}\sum_{n=0}^{N-1}  L^A_nL^A_n \,, \qquad
   V = - \frac{i}{2} \sum_{n=0}^{N-1}\chi_n^A U_n^{AB} \chi_{n+1}^B \,.
 }
We will expand the eigenvalues and eigenstates of $h$ around $x = 0$.

The unperturbed states are eigenstates of $h_0$.  Since $h_0$ is just the gauge kinetic term and does not act on the qubit factor in ${\cal H} = {\cal H}_\text{qubits} \otimes {\cal H}'$, the eigenstates will be at least $\dim {\cal H}_\text{qubits} = 2^{\frac N2}$-fold degenerate. The lowest three energy levels $\epsilon_m$ (defined by $h_0 \ket{\psi_m} = \epsilon_m \ket{\psi_m}$) are given in the following table.
\begin{center}
\begin{tabular}{c|c|c|c}
 $m$ & $\psi_m$ & $\epsilon_m$ & degeneracy \\
 \hline
 $0$ & $\ket{ s_0\ldots s_{\frac N2-1}} \otimes \ket{\textstyle\frac{1}{2} 0 \frac{1}{2}0 \ldots\frac{1}{2}0} $ & $\epsilon_0 = \frac{3N}{16}$ & $2^\frac{N}{2}$ \\
 $1$ & $\ket{ s_0\ldots s_{\frac N2-1}} \otimes \ket{\textstyle\frac{1}{2} 0 \ldots \frac{1}{2}0 \frac 12 1 \frac 12 0 \ldots\frac{1}{2}0} $ & $\epsilon_1 = \frac{3N}{16} + 1$ & $N 2^{\frac N2-1}$ \\
 $2$ & $\ket{ s_0\ldots s_{\frac N2-1}} \otimes \ket{\textstyle\frac{1}{2} 0 \ldots \frac{1}{2}0 \frac 12 1 \frac 12 0 \ldots \frac{1}{2}0 \frac 12 1 \frac 12 0 \ldots\frac{1}{2}0} $ & $\epsilon_2 = \frac{3N}{16} + 2$ & $N(N-2) 2^{\frac N2-3}$
\end{tabular}
\end{center}
In particular, the lowest level is obtained by minimizing the $\grSU(2)$ spins $\ell_n$ on the links. The first few excited levels are obtained by replacing some of the instances of $\ell_n = 0$ with $\ell_n = 1$, and each such replacement increases $\epsilon$ by 1.\footnote{The lowest level where we can replace an $\ell_n = 1/2$ with $\ell_n = 3/2$ is the fifth excited level.}

Let us now focus on obtaining corrections to the lowest energy level $\epsilon_0$. To deal with the $2^{\frac N2}$-fold degeneracy, we will apply Brillouin-Wigner perturbation theory \cite{messiah1999quantum} which we will now review in the context of our perturbation problem.

We define the projection operator onto the degenerate subspace of energy $\epsilon_0$ by
\es{GotP}{
	P = \mathds{1} \otimes \ket{\textstyle\frac{1}{2} , 0, \frac{1}{2}, 0 \cdots, \frac{1}{2}, 0}\bra{\textstyle\frac{1}{2}, 0, \frac{1}{2}, 0 \cdots, \frac{1}{2}, 0} \,.
}
The Brillouin-Wigner perturbation theory starts with the observation that the eigenvalue equation $(h_0+xV)\ket{\psi}= \pE\ket{\psi}$ can be used to reconstruct $\ket{\psi}$ from its projection onto the degenerate subspace of energy $\epsilon_0$, provided that we know the eigenvalue $\pE$. The reconstruction is given by\footnote{\eqref{eq:full_state} can be derived as follows.  We first act with $\mathds{1}-P$ on the eigenvalue equation and obtain $(\mathds{1}-P)(\pE - h_0) \ket{\psi} = (\mathds{1}-P) x V \ket{\psi}$.  Making use of the fact that the operators $\mathds{1}-P$ and $\pE - h_0$ commute, we have $(\pE - h_0)  (\mathds{1}-P)\ket{\psi} = (\mathds{1}-P) x V \ket{\psi}$.  Multiplying both sides by $(\pE - h_0)^{-1}$, this gives $  (\mathds{1}-P)\ket{\psi} = x R_{\cal E} V \ket{\psi}$.  Separating out $P\ket{\psi}$ on the LHS, we then have $P\ket{\psi} = (\mathds{1} - x R_{\cal E} V) \ket{\psi}$.  Lastly, multiplying both sides by $(\mathds{1} - x R_{\cal E} V)^{-1}$ we obtain \eqref{eq:full_state}.}  
\begin{align}\label{eq:full_state}
	\ket{\psi}=(\mathds{1}-xR_{\cal E} V)^{-1}P\ket{\psi}, \qquad R_{\cal E} = (\pE -h_0)^{-1}(\mathds{1}-P)\,,
\end{align}
Using this, we can recast the eigenvalue problem as
\begin{align}\label{eq:bw_pert}
	P[\epsilon_0+xV(\mathds{1}-xR_{\cal E} V)^{-1}]P\ket{\psi}=\pE P\ket{\psi} \,.
\end{align}
In this presentation, we can solve for the projection $P\ket{\psi}$ rather than for the full eigenvector $\ket{\psi}$. The eigenvalue $\pE$ appears explicitly on the right and also on the left within $R_{\cal E}$, so we must solve perturbatively in $x$ for both $\pE$ and $P\ket{\psi}$.  Once $\pE$ and $P\ket{\psi}$ are known to the desired order in $x$, one can recover the eigenstate in the full Hilbert space by applying \eqref{eq:full_state}.  Depending on the details of the perturbation $V$, the initial degeneracy may be partially or fully resolved at higher orders in $x$.

To proceed, we consider the power series ansatz
\begin{align}\label{eq:expansions}
	\pE=\pE^{(0)}+x\pE^{(1)}+x^2 \pE^{(2)}+\cdots\,, \qquad P\ket{\psi} = \ket{\psi^{(0)}}+x\ket{\psi^{(1)}}+x^2\ket{\psi^{(2)}}+\cdots \,.
\end{align}
The projected operator in the eigenvalue equation \eqref{eq:bw_pert} makes no reference to the gauge field configuration and can be viewed as an effective Hamiltonian for the factor ${\cal H}_\text{qubits}$ of the full Hilbert space that describes the $N/2$ qubits.  Indeed, we can write
 \es{LHSExpansion}{
	P[\epsilon_0+xV(\mathds{1}-xRV)^{-1}]P 
	 =  P\left[\epsilon_0+xh^{(1)}+x^2h^{(2)}+x^3h^{(3)}+x^4 h^{(4)}+ {\cal O}(x^5)\right]\otimes \mathds{1} \,,
 }
with the effective Hamiltonian given in the square bracket of \eqref{LHSExpansion}, and 
 \es{hDefs}{
  P(h^{(1)}\otimes \mathds{1}) = PVP \,, \qquad P(h^{(2)}\otimes \mathds{1}) = PV\frac{\mathds{1}-P}{\pE-h_0}VP \,, \qquad \text{etc.}
 }
In \eqref{hDefs}, we postpone expanding $\pE$ as in \eqref{eq:expansions}.

At first order, we need to determine $PVP$ by projecting onto the ground states, acting with $V$, and then projecting again onto the ground states. The projectors $P$ and $\mathds{1}-P$ both act on ${\cal H}_\text{qubits}$ as the identity, so the nontrivial action on the qubits comes only from $V$. Let us act with $PVP$ on the state $\ket{\chi}\otimes \ket{\textstyle\frac{1}{2} 0 \frac{1}{2}0 \ldots\frac{1}{2}0}$. Using the rules explained in Section~\ref{GAUGEINV}, the actions of the terms of the hopping operator are
\begin{subequations}
	\begin{align}
	\left(-\frac{i}{2}\chi^A_{2k}U^{AB}_{2k}\chi^B_{2k+1}\right)\ket{\chi}\otimes\ket{\ldots 0\overbracket{\textstyle\frac{1}{2}}^{\ell_{2k}} 0 \ldots} &= -\frac{3}{4}(Z_k\ket{\chi})\otimes\ket{\ldots\textstyle0 \frac{1}{2} 0 \ldots} \,, \label{eq:even_action}\\
	\left(-\frac{i}{2}\chi^A_{2k+1}U^{AB}_{2k+1}\chi^B_{2k+2}\right)\mathds{1}\otimes\ket{\ldots \textstyle\frac{1}{2}\overbracket{0}^{\ell_{2k+1}} \textstyle\frac{1}{2} \ldots} &= \frac{\sqrt{3}}{4}(X_kX_{k+1}\ket{\chi})\otimes \ket{\ldots\textstyle \frac{1}{2} 1 \frac{1}{2} \ldots} \,.\label{eq:odd_action}
\end{align}
\end{subequations}
For the first order term, $PVP$, the second factor of $P$ will annihilate the right hand side of \eqref{eq:odd_action}, so only \eqref{eq:even_action} contributes, and we find
\begin{align}\label{eq:first_correction}
	h^{(1)} = -\frac{3}{4}\sum_{k=0} ^{N/2-1} Z_k \,.
\end{align}

We can now proceed to compute the second order term $ PV\frac{\mathds{1}-P}{\pE-h_0}VP$. This time, after acting with $VP$, it is only the right hand side of \eqref{eq:odd_action} that survives the action of $\mathds{1}-P$. We then have to get back to the ground state by applying the perturbation $V$ again, which means we have to act with the same odd hopping operator. From
\begin{equation}
\begin{split}
	&\left(-\frac{i}{2}\chi^A_{2k+1}U^{AB}_{2k+1}\chi^B_{2k+2}\right)\frac{\mathds{1}-P}{\pE-h_0}\left(-\frac{i}{2}\chi^A_{2k+1}U^{AB}_{2k+1}\chi^B_{2k+2}\right)\ket{\chi}\otimes\ket{\ldots \textstyle\frac{1}{2}{0} \textstyle\frac{1}{2} \ldots}\\
	 &\qquad= \frac{3}{16}\frac{1}{\pE-\epsilon_1}\ket{\chi}\otimes \ket{\ldots\textstyle \frac{1}{2} 0 \frac{1}{2} \ldots} \,, 
\end{split}
\end{equation}
we find in total
\begin{align}\label{eq:2nd_corr}
	h^{(2)}  = \frac{3N}{32}\frac{1}{\pE-\epsilon_1}\mathds{1} \,.
\end{align}

We can continue this program to higher orders.  The third and fourth order terms are
\begin{align}
h^{(3)}&=\left[-\frac{3(3N-8)}{128}\sum_{k=0}^{N/2-1}Z_k-\frac{3}{32}\sum_{k=0}^{N/2-2}(-{\cal F})^{\delta_{k, \frac{N}{2}-1}}X_kX_{k+1}\right]\frac{1}{(E-\epsilon_1)^2} \,, \\
	h^{(4)}&=\left[\frac{9}{64}\sum_{k=0}^{N/2-2}X_kX_{k+1}\left(\sum_{k'\neq k,k+1} Z_{k'}\right)-\frac{9}{64}\mathcal F X_{N/2-1}X_0\left(\sum_{k\neq N/2-1,0} Z_k\right)\right]\frac{1}{(E-\epsilon_1)^3}\nonumber\\
	&{}+ \frac{9N(N-2)}{512}\frac{1}{(E-\epsilon_1)^2(E-\epsilon_2)}+\frac{9N}{128}\frac{1}{(E-\epsilon_1)^3}\\
	&{}+\frac{9(3N-16)}{512}\left(\sum_{k=0}^{N/2-1} Z_k\right)^2\frac{1}{(E-\epsilon_1)^3} +\frac{3}{32}\sum_{k=0}^{N/2-1} Z_k Z_{k+1}\frac{1}{(E-\epsilon_1)^3} \,.\nonumber
\end{align}
Note that the first-order term breaks the degeneracy of the ground state, but still leaves the first excited state $\frac{N}{2}$-fold degenerate. The second-order term does not break any degeneracies, and so it is only at third-order when the first excited state becomes unique. Using \eqref{eq:bw_pert}, \eqref{eq:expansions}, \eqref{LHSExpansion}, and the expressions for $h^{(i)}$ above we can solve for the two lowest eigenvalues up to fourth order:
\begin{subequations}\label{eq:strong_energy}
	\begin{align}
	\pE_0&= \frac{3 N}{16}-\frac{3   N}{8}x-\frac{3  N}{32}x ^2+\frac{3 
		N}{32}x ^3-\frac{51  N}{512}x ^4+\mathcal O(x ^5) \,, \\
	\pE_1 &= \frac{3 N}{16}-\frac{3}{8} x  (N-4)-\frac{3  N}{32}x ^2+\frac{3(N-6)}{32} 
		x ^3-\frac{3(17 N-64)}{512} x ^4 +\mathcal O(x ^5) \,.
\end{align}
\end{subequations}
The bare eigenvalues are extensive in the system size, but the gap to the lowest excitation is an intensive quantity
\begin{align}\label{eq:pade}
	\delta \pE \equiv \pE_1 - \pE_0 = \frac{3  }{2}x-\frac{9 }{16}x ^3+\frac{3 }{8}x ^4+\mathcal O(x ^5) \,.
\end{align}

We can now extrapolate this result derived in the strong coupling limit $x\ll 1$ to the continuum limit $x\gg 1$ using a Pad\'e approximant. From \eqref{eq:strong_setup}, we see that the continuum limit of the gap $E_1 - E_0 = \frac{g_\text{YM}}{\sqrt{x}} \delta \pE$ is finite only when $\delta \pE$ scales as $\sim \sqrt{x}$. This constraint allows for several different approximation schemes. We found the most accurate result by applying a $(0,2)$ Pad\'e approximant to $\left(\frac{\delta\pE}{x}\right)^4$, giving
 \es{Pade}{
	\delta \pE = x\left(\left(\frac{\delta\pE}{x}\right)^4\right)^{1/4} = \frac{3x}{2}\left(\frac{1}{1+\frac{3x^2}{2}}\right)^{1/4}+\mathcal O(x^4) \,.
 }
This gives a continuum estimate for the energy gap of
\begin{align}
	E_1-E_0 \approx \gym \left(\frac{3}{2}\right)^{3/4} \approx 1.355 \, \gym \,.
\end{align}
This result agrees well with the lattice results of Section \ref{NUMERICS}, along with the results of DLCQ given in \eqref{eq:dlcq}.

We can also estimate the vacuum expectation value of the fermion bilinear operator, $\braket{\tr \bar\psi \psi}$, using the strong-coupling expansion. 
In the continuum, we have
\begin{align}
\braket{\tr \bar\psi \psi} = \frac{1}{L} \left.\frac{\partial E_0}{\partial m}\right\vert_{m=0}
\end{align}
On the lattice, this can be computed from the expectation value of the mass operator using the identification
\begin{align}\label{eq:mass_id}
	\braket{\tr \bar\psi \psi} = \frac{1}{aN}\braket{0|H_\text{mass}|0},\qquad H_\text{mass}= - \frac{i}{2}\sum_{n=0}^{N-1}   (-1)^n \chi_n^A U_n^{AB}  \chi_{n+1}^B  \,.
\end{align}
To compute the expectation value $\braket{0|H_\text{mass}|0}$, we need the ground state in the full Hilbert space, which can be computed using \eqref{eq:full_state}.   To third order, we find
\begin{align}\label{eq:condensate_sce}
	N^{-1}\braket{0|H_\text{mass}|0} = \frac{3}{8}-\frac{3}{16}x+\frac{3}{32}x^2-\frac{3}{128}x^3 +\mathcal O(x^4) \,.
\end{align}
The identification \eqref{eq:mass_id} requires that $N^{-1}\braket{0|H_\text{mass}|0}\sim \sqrt{x}$ at large $x$ if we are to have a finite continuum limit. Hence, it is appropriate to approximate the sixth power of the right hand side of \eqref{eq:condensate_sce} using a $(0,3)$ Pad\'e approximant.  Such an approximation gives an estimate of $\braket{\tr \bar\psi \psi} \approx-0.33\gym$.

\section{Numerical results for $\grSU(2)$}
\label{NUMERICS}

Here we use the formulation of Section \ref{GAUGEINV} to explicitly calculate the low-lying spectrum of the lattice Hamiltonian for the $\grSU(2)$ theory. In Section \ref{sec:large_mass}, we show that we recover the expected energies for a pure $\grSU(2)$ gauge theory on a circle when the adjoint fermion mass is made large. In Section \ref{sec:massless}, we study the spectrum of the massless theory in detail, showing good agreement with results obtained from DLCQ \cite{Antonuccio:1998uz,Dempsey:2022uie} and with the strong coupling expansions given in Section \ref{APPROXIMATIONS}. Finally, in Section \ref{sec:massive} we turn on the adjoint mass and show further agreement with DLCQ in the trivial universe, along with new results for the nontrivial universe.

For all the exact diagonalization results shown in this section, we use PETSc and SLEPc \cite{petsc-user-ref,petsc-efficient,slepc-toms,slepc-manual}.

\subsection{Large mass limit}\label{sec:large_mass}

In the continuum theory in the $m\to \infty$ limit, we expect the adjoint fermion to decouple, leaving behind a pure $\grSU(2)$ gauge theory. The energy levels for this theory on a circle of length $L$ are given by
\begin{equation}\label{eq:pure_su2}
	E_\ell = \frac{g_\text{YM}^2 L}{2} \ell(\ell+1).
\end{equation}
In the $p = 0$ universe we have $\ell = 0, 1, \ldots$, and in the $p = 1$ universe we have $\ell = \frac{1}{2}, \frac{3}{2}, \ldots$. Therefore, the energy gaps above the vacuum will be
\begin{subequations}\label{eq:decoupled_gaps}
	\begin{align}
		\Delta E_n &= E_n - E_0 = \frac{g_\text{YM}^2 L}{2} n(n+1) \qquad \text{($p = 0$ universe),}\label{eq:trivial_gaps}\\
		\Delta E_n &= E_{n + \frac 12} - E_{\frac 12} = \frac{g_\text{YM}^2 L}{2} n(n + 2) \qquad \text{($p = 1$ universe),}\label{eq:nontrivial_gaps}
	\end{align}
\end{subequations}
where $n = 0, 1, 2, \ldots$.

\begin{figure}[t]
	\centering
	\includegraphics[width=.7\linewidth]{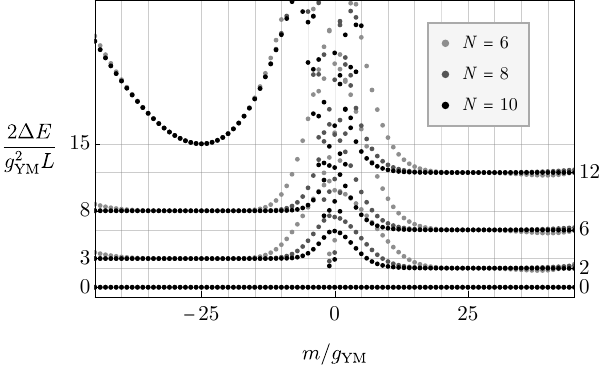}
	\caption{For the adjoint mass in a range of large values, the lowest energy gaps reproduce those of a pure $\grSU(2)$ gauge theory, given in \eqref{eq:decoupled_gaps}. The lattice spacing is fixed to $\gym a = 0.04$, and the energy gaps are normalized by $\frac{1}{2}g_\text{YM}^2 L$ so we can compare lattices of sizes $N = 6, 8$, and 10. The maximum link representation is $\ell_\text{max} = 4$, which allows us to see continuum energy levels corresponding to representations up to spin $\frac{7}{2}$. When $|m| = a^{-1}$, we can understand this behavior analytically as coming from alternating links of the lattice, with the types of representations appearing on the links determined by the sign of $m$ (see Appendix \ref{app:large_mass}).}
	\label{fig:decoupled}
\end{figure}

As explained in Section~\ref{CONSEQUENCES}, the two universes of the $N_c=2$ theory are connected by the chiral symmetry transform $\mathcal V$ which acts on the Hamiltonian by flipping the sign of the mass. If we restrict to the $p = 0$ universe and take $m\rightarrow \infty$ then one will recover the trivial spectrum \eqref{eq:trivial_gaps}. On the other hand the chiral transformation $\mathcal V$ ensures that the limit $m\rightarrow -\infty$ will result in the non-trivial spectrum \eqref{eq:nontrivial_gaps}.

In Figure \ref{fig:decoupled}, we show that this behavior is reproduced in our lattice theory. We take the lattice spacing to be $g_\text{YM} a = 0.04$ and use lattice sizes of $N = 6, 8, 10$, for a large range of masses. We find that the energy gaps precisely reproduce the sequences in \eqref{eq:decoupled_gaps} up to the level with spin $\frac{7}{2}$, which is related to the link representation cutoff of $\ell_\text{max} = 4$. For $m > 0$ we have the sequence for the trivial flux tube sector, and for $m < 0$ we have the sequence for the nontrivial flux tube sector.

Note that the spin $\frac{7}{2}$ level only has the correct continuum energy at $m = -25\gym = -1/a$. In fact, we can understand all of these states at $|m| = a^{-1}$ via perturbation theory on the lattice. This is discussed in Appendix \ref{app:large_mass}. Numerically, we find that the convergence away from $|m| = a^{-1}$ improves rapidly with increasing $N$ and $\ell_\text{max}$.

\subsection{Massless theory}\label{sec:massless}

\begin{figure}
	\centering
	\includegraphics[width=.8\linewidth]{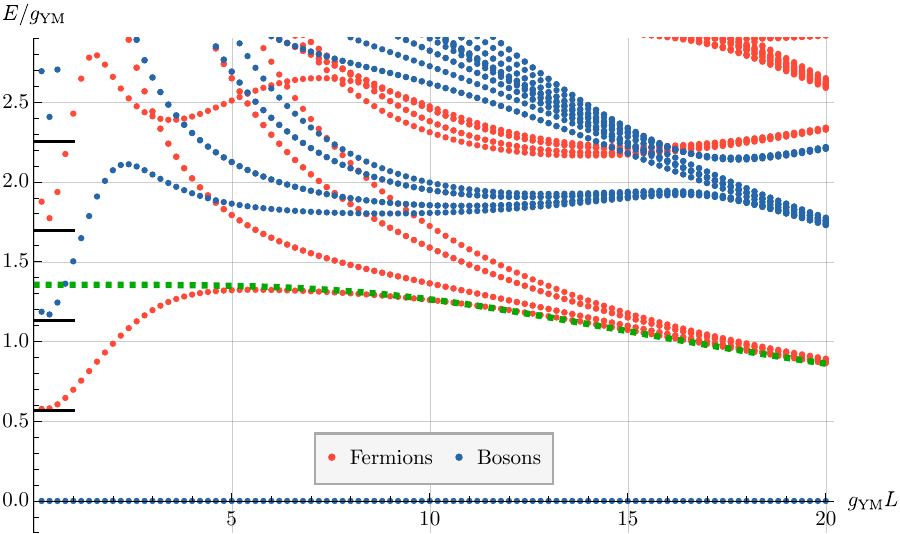}
	\caption{The low-lying spectrum of the lattice Hamiltonian for $m=0$ as a function of $\gym L$, for $N = 12$ sites with periodic boundary conditions. The spectrum on a small circle consists of equally spaced levels of a harmonic oscillator with frequency $\omega = \frac{g}{\sqrt{\pi}}$, and these levels are marked with black lines. The Pad\'e approximant \eqref{Pade} to the strong coupling expansion for the lowest fermion mass is plotted in green.}
	\label{fig:massless_spectrum}
\end{figure}

We will now set $m = 0$ and aim to study the spectrum of the $\grSU(2)$ gauge theory on a circle of length $L=Na$. In Figure~\ref{fig:massless_spectrum}, we give the spectrum for $N = 12$ sites as a function of the circle length $\gym L$, for the $p=0$ universe.  (The spectrum of the $p=1$ universe is identical to that of the $p=0$ universe, with bosons and fermions swapped.) We find that the low-lying spectrum is fairly well-converged in $N$ even on this small lattice, so that Figure~\ref{fig:massless_spectrum} is representative of the large $N$ spectrum.

The spectrum in Figure~\ref{fig:massless_spectrum} can be understood analytically at small and large $\gym L$.  First, in the limit $\gym L \ll 1$, the dynamics reduces to that of the gauge holonomy.  In the continuum theory such an analysis was performed in \cite{Lenz:1994du} in the case where the fermions obey anti-periodic boundary conditions and in \cite{Cherman:2019hbq} in the case of periodic boundary conditions.  In the case of periodic boundary conditions, one finds the spectrum of a harmonic oscillator with frequency $\frac{1}{\sqrt{\pi}}\gym$.  This equally-spaced harmonic oscillator energy levels are marked with black lines in Figure~\ref{fig:massless_spectrum}.

\begin{figure}[t]
	\centering
	\begin{subfigure}[t]{\linewidth}%
	\centering
	\includegraphics[width=.55\linewidth]{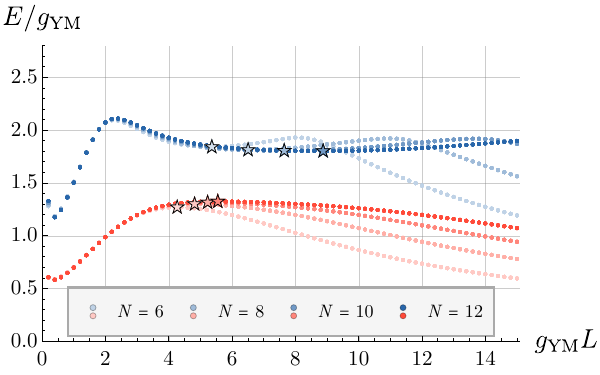}
	\caption{For a given lattice size $N$, we estimate the mass of the lowest fermion and boson for the $p = 0$ universe of the $m = 0$ theory using local extrema of their energies as functions of the circle length. These extrema are marked with stars.}
	\label{fig:massless_plateaus}
	\end{subfigure}\\[1em]
	\begin{subfigure}[t]{\linewidth}%
	\centering
	\includegraphics[width=.55\linewidth]{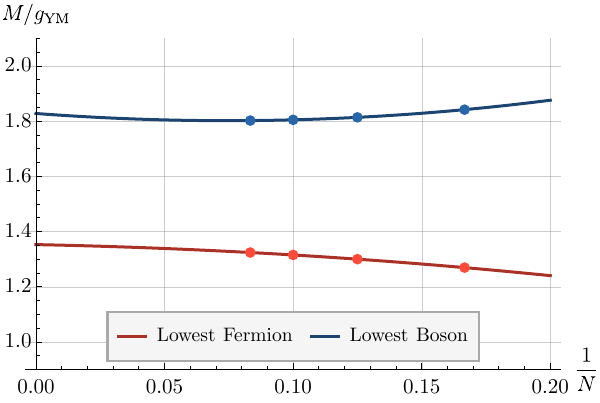}
	\caption{The masses of the lowest fermion and boson for the $p = 0$ universe of the $m = 0$ theory extrapolated to $N\to\infty$ are as in \eqref{eq:lowest_fermion} and \eqref{eq:lowest_boson} respectively.}
	\label{fig:massless_extrapolation}
	\end{subfigure}
	\caption{}
\end{figure}

In the limit $\gym L  \gg 1$, for fixed $N$, we can use the strong coupling expansion developed in Section \ref{APPROXIMATIONS}.  The Pad\'e approximant \eqref{Pade}, gives the gap estimate
 \es{GapEst}{
  E_1 - E_0 \approx \gym \frac{3}{2}\left(\frac{1}{\frac{3}{2} + \gym^2 a^2}\right)^{1/4} \,,
 }
which provides a very good approximation for the gap for all $\gym a \gtrsim 0.5$.  In Figure~\ref{fig:massless_spectrum}, we plot the estimate \eqref{GapEst} using a dashed green line.

Between the small-circle and strong-coupling regimes, we see that the lowest fermionic and bosonic excitations start developing approximate plateau regions.  We found that these plateau regions extend to larger and larger values of $\gym L$ as we increase $N$, as shown in Figure~\ref{fig:massless_plateaus}, and we believe that it is these regions that we should extrapolate to large $N$ in order to extract the infinite-volume continuum spectrum.  For the lowest fermionic excitation, a proxy for where the plateau occurs is the local maximum, while for the lowest bosonic excitation, the analogous proxy would be a local minimum.  The series of these local maxima and minima for $N=6, 8, 10, 12$ are marked with stars in Figure~\ref{fig:massless_plateaus}.   Extrapolating, we find
\begin{equation}\label{eq:lowest_fermion}
	M_f/\gym \approx 1.35
\end{equation}
for the lowest fermion bound state mass in the continuum limit, and 
\begin{equation}\label{eq:lowest_boson}
	M_b/\gym \approx 1.83
\end{equation}
for the lowest bosonic excitation (see Figure~\ref{fig:massless_extrapolation}). The lowest fermion and boson masses were also calculated using DLCQ in \cite{Dempsey:2022uie}, and in these units were
\begin{equation}\label{eq:dlcq}
	M_f^\text{(DLCQ)}/\gym \approx 1.35\,, \qquad M_b^\text{(DLCQ)}/\gym \approx 1.85 \,,
\end{equation}
so the two methods agree well.

\begin{figure}
	\centering
	\includegraphics[width=.8\linewidth]{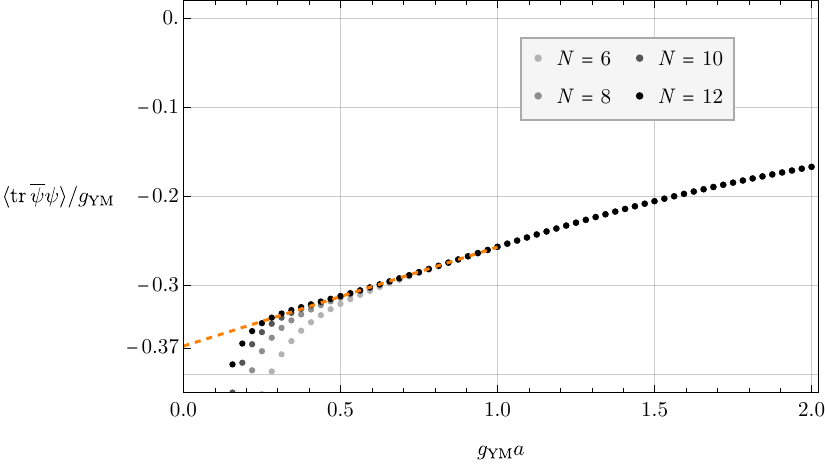}
	\caption{The vacuum expectation value of the mass operator $\tr \bar\psi \psi$ in the $p = 0$ universe as a function of the lattice spacing for $N = 6,8,10$, and 12. By extrapolating to $\gym a \to 0$ from the region converged in $N$, we obtain a continuum estimate of $\langle \tr \bar\psi \psi\rangle/\gym \approx -0.37$.}
	\label{fig:chiral_condensate}
\end{figure}

One advantage of the equal-time quantization is that we have access to the vacuum state and its properties. As an example, in Figure \ref{fig:chiral_condensate} we plot the chiral condensate $\langle \tr \bar\psi \psi\rangle$ in the $p = 0$ universe as a function of the lattice spacing for several values of $N$\@. Extrapolating to $a\to 0$ from the region where the results are converged in $N$, we obtain a continuum estimate of $\langle \tr \bar\psi \psi\rangle/\gym \approx -0.37$. For the $p = 1$ universe, we would find the opposite sign. This is in relatively good agreement with the value of $-0.33$ obtained from the strong coupling expansion in Section \ref{APPROXIMATIONS}.

\subsection{Massive theory}\label{sec:massive}

\begin{figure}
	\centering
	\includegraphics[width=.8\linewidth]{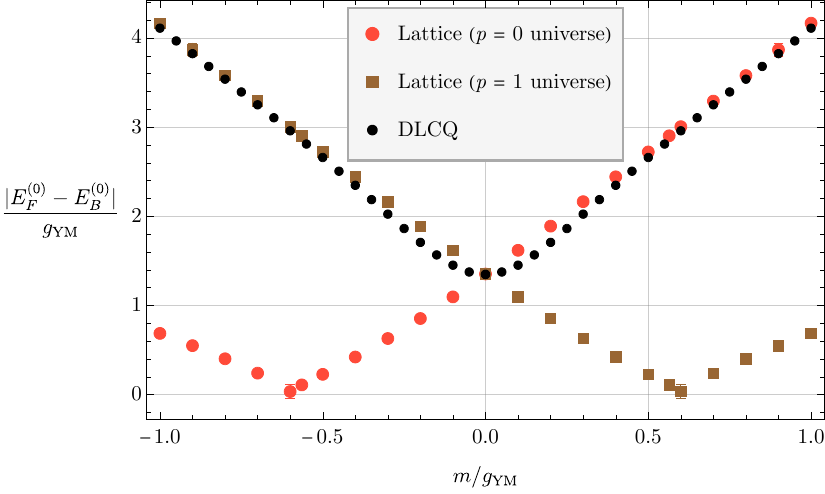}
	\caption{For adjoint masses $-\gym \le m\le \gym$, we compute the energy gap in both universes using the same method as for the massless theory. For $m\ge 0$ in the $p = 0$ universe, or $m\le 0$ in the $p = 1$ universe, we can compare this with the lowest fermion mass computed using DLCQ in \cite{Dempsey:2022uie}. We obtain excellent agreement,
except at small $(m/\gym)^2$ where the mass dependence is known to converge slowly in DLCQ.
 We also see the gap closing near $m = m_\text{SUSY}$ in the $p = 1$ universe, or $m = -m_\text{SUSY}$ in the $p = 0$ universe.}
	\label{fig:fermion_mass}
\end{figure}

Once we turn on a mass $m$ for the adjoint fermion, we can continue to extract the lowest excitations above the vacuum by extrapolating plateaus like those of Figure~\ref{fig:massless_plateaus} to $N\to\infty$. In Figure \ref{fig:fermion_mass}, we plot the mass of the lightest particle with fermion parity opposite that of the vacuum,\footnote{With our relatively small lattices, we do not yet have sufficient precision to extract the lightest particle with fermion parity equal to that of the vacuum.} extracted from the lattice spectra as a function of the adjoint mass for both universes.  For $m\geq 0$ in the $p = 0$ universe, we see that there is good agreement with DLCQ at $m = 0$ and at sufficiently large mass. For $m\le 0$, DLCQ instead agrees with the results from the $p = 1$ universe. The discrepancy in either case near $m = 0$ can be attributed to the light-cone Hamiltonian only depending on $m^2$. Thus, when DLCQ is not completely converged to its continuum limit, it will struggle to capture the linear behavior near $m = 0$ that is related to the non-vanishing expectation values of the mass operator.

We see in Figure~\ref{fig:fermion_mass} that the gap in the $p = 1$ universe closes at $m/\gym = \pi^{-1/2} \approx 0.56$, as expected from the presence of the massless Goldstino of spontaneously broken supersymmetry \cite{Dubovsky:2018dlk}.\footnote{The discontinuity of the derivative is because the ground state becomes fermionic for
$m/\gym > \pi^{-1/2}$. Had we plotted the energy difference between the lowest fermionic and bosonic states, there would be no such discontinuity.} 
 Likewise, the gap closes in the $p = 0$ universe at $m/\gym = -\pi^{-1/2}$. At these points, the IR sector should be described by one free massless Majorana fermion on the worldsheet of the confining flux tube.

\section{Discussion}
\label{DISCUSSION}

In this work, we introduced a new Hamiltonian lattice gauge theory model for adjoint QCD$_2$ with gauge group $\grSU(N_c)$.  As we explained in Section~\ref{LATTICE}, this lattice model uses staggered Majorana fermions.
The resulting lattice Hamiltonian is given by \eqref{eq:lattice_hamiltonian}, and it should be supplemented by the Gauss law constraint \eqref{eq:gauss_law}. 

In Section~\ref{SYMMETRIES}, we analyzed the symmetries and the anomalies of this lattice model.  We found lattice analogs of the $\Z_2^{[1]} \rtimes (\Z_2)_C \times (\Z_2)_F$ symmetries present for all $m$, and, for $m=0$, we also found a  lattice analog of the $(\Z_2)_\chi$ axial symmetry, which is represented by translation by one lattice site.  Interestingly, the lattice model exhibits analogs of the mixed 't Hooft anomalies of the continuum theory, and these mixed anomalies have various implications on the spectrum both for $m=0$ and for $m \neq 0$. 
In recent literature, there have been interesting investigations of lattice non-invertible symmetries \cite{Cheng:2022sgb,Seiberg:2023cdc}.
We leave further related studies of our lattice model for future work. 

In Sections~\ref{APPROXIMATIONS} and \ref{NUMERICS}, we then studied the $N_c = 2$ theory in more detail, both analytically in the strong coupling expansion and numerically using exact diagonalization.  For $m \geq 0$ in the $p = 0$ universe, or for $m\leq 0$ in the $p = 1$ universe, our results are in good agreement with the DLCQ spectrum computed in \cite{Dempsey:2022uie}.  The $m>0$ spectrum that we find in the $p = 1$ universe is new, and it would be interesting to reproduce it using DLCQ\@.  Its main feature is that the gap closes at $m = m_\text{SUSY}$ as expected from the appearance of a massless particle due to spontaneous supersymmetry breaking.  As explained in Section~\ref{SYMMETRIES}, the spectrum of the $p = 1$ universe for given $m$ matches that of the $p = 0$ universe at $-m$ (with bosons and fermions swapped).

Stepping back, there has been renewed interest in the Hamiltonian formulation of lattice gauge theory. One of the reasons is that, for theories in $1+1$ dimensions, it is possible to apply tensor network methods to achieve high numerical precision. Another reason is the possibility of quantum simulations using specially designed experimental devices. Such simulators have been constructed for the lattice one-flavor Schwinger model using, for example, trapped ions \cite{Kokail:2018eiw}. 
This model contains one Dirac, or equivalently two Majorana, fermion degrees of freedom per lattice site.  The next-simplest model appears to be adjoint QCD$_2$ with gauge group $\grSU(2)$, whose lattice implementation we constructed in this paper. This model is also gapped and contains only three Majorana degrees of freedom per lattice site. While in this paper we carried out exact diagonalizations of this model on a periodic chain, it would be interesting to also study our model using tensor network methods.  Such a study would hopefully lead to significantly better numerical precision.  We also hope that the relative simplicity of the $\grSU(2)$ model can eventually allow an experimental construction of a quantum simulator.

Furthermore, it is highly desirable to carry out a numerical study of our lattice model with gauge group $\grSU(3)$. Here, in the continuum treatment of the massless theory, there are $4$ topological sectors \cite{Cherman:2019hbq,Komargodski:2020mxz}. Two of them have zero triality;
in our lattice model they are related by a lattice translation by one site.
 The other two sectors have trialities $1$ and $2$; they correspond to a flux tube wound around the circle in the two possible directions. On a lattice, the ground states in these sectors are not degenerate with the ground state of the zero triality sector. Therefore, unlike in the $\grSU(2)$ model, the vanishing of the string tension in the $\grSU(3)$ theory with a massless adoint Majorana fermion
\cite{Gross:1995bp,Komargodski:2020mxz,Dempsey:2021xpf}
is not automatic in our lattice model: it may require a careful numerical extrapolation to the large volume limit and perhaps a fine tuning of the $4$-fermion operators in the lattice Hamiltonian. In view of having $8$ Majorana fermions per lattice site, this model is certainly challenging numerically, but we hope to carry out its initial studies in the future.

\section*{Acknowledgments}

We are grateful to Bernardo Zan for collaboration at the early stages of this project. We also thank O. Aharony, A. Cherman and D. Gaiotto for useful discussions.  This work was supported in part by the US National Science Foundation under Grants No.~PHY-2111977 and PHY-2209997, and by the Simons Foundation Grants 
No.~488653 and 917464.  RD was also supported in part by an NSF Graduate Research Fellowship.

\appendix

\section{Quantum particle moving on a group manifold}
\label{GROUPMANIFOLD}

\subsection{Lagrangian and Hamiltonian}

Consider a quantum particle moving on a group manifold $G$, with Lagrangian
 \es{LagParticle}{
  {\cal L} = -\frac{1}{g^2} \tr \left[ \dot U U^{-1} \dot U U^{-1} \right] \,,
 }
 where $U$ is a group element.
  Suppose we parameterize $U$ using angular coordinates $\theta^\mu$.  Then $i (\partial_\mu U) U^{-1}$ is a Lie-algebra-valued vector field on the group manifold, so it can be written as a linear combination of the Lie algebra generators $T^A$, with coefficients $e_\mu^A$ being frame vectors on the group manifold:
  \es{vDef}{
   i (\partial_\mu U) U^{-1} = e_\mu^A T^A \,,  \qquad 
     e_\mu^A = 2 \tr \left[ T^A i (\partial_\mu U) U^{-1} \right] \,.
  }
Similarly, $i U^{-1} \partial_\mu U$ is also a linear combination of the Lie algebra generators, with different coefficients $f_\mu^A$, which form another set of frame vectors on the group manifold:
  \es{fDef}{
   i U^{-1} (\partial_\mu U)  = f_\mu^A T^A \,,  \qquad 
     f_\mu^A = 2 \tr \left[ T^A i  U^{-1} (\partial_\mu U) \right] \,.
  }
With the normalization $\tr (T^A T^B) = \delta^{AB}/2$, one can then show that the two sets of frame vectors are related to each other via the matrix $U^{AB} \equiv 2 \tr (T^A U T^B U^{-1})$, in particular
 \es{FrameRelations}{
  e_\mu^A = U^{AB} f_\mu^B \,, \qquad f_\mu^A = U^{BA} e_\mu^B \,.
 }

We can rewrite the Lagrangian in terms of the angles $\theta^\mu$ as
 \es{LagParticle2}{
  {\cal L} = \frac{1}{2 g^2} e_\mu^A \dot \theta^\mu e_\nu^A \dot \theta^\nu
   = \frac{1}{2 g^2} f_\mu^A \dot \theta^\mu f_\nu^A \dot \theta^\nu   \,.
 }  
As usual, we can define the canonical momentum  (the minus sign is for later convenience) $-\pi_\mu = \frac{\partial L}{\partial \dot \theta^\mu} = \frac{1}{g^2} e_\mu^A e_\nu^A \dot \theta^\nu = \frac{1}{g^2} f_\mu^A f_\nu^A \dot \theta^\nu$.  The left-acting/right-acting canonical momentum operators (which in our lattice setup become the left-acting/right-acting electric fields) are defined by contracting $\pi_\mu$ with the corresponding inverse frame vectors.  In particular, we write $\pi_\mu = e_\mu^A L^A = f_\mu^A R^A$, from which we can solve for $L^A$ and $R^A$:
 \es{pitoL}{
   L^A = \tilde e^{\mu A} \pi_\mu \,, \qquad R^A = \tilde f^{\mu A} \pi_\mu \,,
 }
 where $\tilde e^{\mu A}$ and $\tilde f^{\mu A}$ are the inverse frames obeying $\tilde e^{\mu A} e_\mu^B = \delta^{AB}$ and $\tilde f^{\mu A} f_\mu^B = \delta^{AB}$, respectively.  Because of \eqref{FrameRelations}, we also have 
   \es{RelLR}{
  L^A = U^{AB} R^B \,, \qquad R^A = U^{BA} L^B \,,
 }
 
 The Hamiltonian is then given by the standard formula $H = \frac{\partial {\cal L}}{\partial \dot \theta^\mu}  \dot \theta^\mu - {\cal L}$, which gives
  \es{GetHamilt}{
    H =  \frac{g^2}{2} L^A L^A =   \frac{g^2}{2} R^A R^A  \,.
  }
 Canonical quantization requires $[-\pi_\mu, \theta^\nu] = \frac{1}{i} \delta_\mu^\nu$.  Using \eqref{pitoL}, this implies the relations
 \es{LU}{
   [L^A, U] = T^A U \,, \qquad [R^A, U] =  U T^A \,.
 }

 \subsection{Properties of the operators $L^A$, $R^A$, and $U^{AB}$}
 
 From \eqref{LU}, the definition $U^{AB} = 2 \tr (T^A U T^B U^{-1})$, as well as the algebra obeyed by the generator
  \es{AlgGen}{
   [T^A, T^B] = i f^{ABC} T^C \,,
  }
we can derive the various commutation relations
  \es{LRUCommut}{
   [L^A, L^B] &= - i f^{ABC} L^C \,, \qquad
   \ \ \ \  [R^A, R^B] = i f^{ABC} R^C \,, \\
   [L^A, U^{BC}] &= - i f^{ABD} U^{DC} \,, \qquad
   [R^A, U^{BC}] = i f^{ACD} U^{BD} \,.
  }
Note that while $R^A$ obey the same algebra as the generators, the $L^A$ obey it with an extra minus sign.

For the operators $U^{AB}$ we also have the relations 
 \es{URelations}{
  U^{AC} U^{BC} &= \delta^{AB} = U^{CA} U^{CB} \,, \qquad
  f^{CDE} U^{AD} U^{BE} = f^{ABF} U^{FC} \,.
 }
 These relations follow directly from the definition $U^{AB} = 2 \tr (T^A U T^B U^{-1})$ as well as the completeness relation $X^A = 2 \tr (X T^A)$ for any quantity in the adjoint representation.
 
 Using \eqref{URelations}, one can show that $L^A = U^{AB}R^B$ is consistent with the commutation relations  \eqref{LRUCommut}, and moreover that the left-acting and right-acting generators commute:
  \es{RLCommut}{
   [L^A, R^B] = 0 \,.
  }
 
 \subsection{Hilbert space for $G = \grSU(2)$}
 
For a general group $G$, the Hilbert space consists of normalizable functions on the group manifold, which, by the Peter-Weyl theorem, can be identified with the space of matrix elements in all irreducible representations of the group.  Let us focus on $G = \grSU(2)$, where these are normalizable functions on the three-sphere.  The Hilbert space splits as
 \es{calH}{
  {\cal H} = \bigoplus_{\ell \in \frac{\Z_+}{2}} V^{(\ell)}_L \otimes V^{(\ell)}_R  \,,
 }
where $V^{(\ell)}$ denotes the $2 \ell + 1$-dimensional vector space corresponding to the spin-$\ell$ representation.  

To construct the states explicitly, let $\ket{\ell, \mathfrak{m}}$ be the standard basis of simultaneous orthonormal eigenstates of $T^A T^A$ and $T^3$:
 \es{estatesT}{
  T^A T^A \ket{\ell, \mathfrak{m}} = \ell(\ell+1) \ket{\ell, \mathfrak{m}} \,, \qquad
   T^3 \ket{\ell, \mathfrak{m}} = \mathfrak{m} \ket{\ell, \mathfrak{m}} \,.
 }
Note that for a general generator $T^A$, which in the spin $\ell$ representation can be written as a matrix $T^A_{\mathfrak{m} \mathfrak{m}'}$, we have\footnote{If we write a general state in the spin-$\ell$ representation as $\psi = a_\mathfrak{m} \ket{\ell, \mathfrak{m}}$, then $T^A \psi = T^A_{\mathfrak{m} \mathfrak{m}'}a_{\mathfrak{m}'} \ket{\ell, \mathfrak{m}}$.  Note that the ordering of the indices of $T^A$ is swapped compared to \eqref{TAction}.}
 \es{TAction}{
  T^A \ket{\ell, \mathfrak{m}} = T^A_{\mathfrak{m}' \mathfrak{m}} \ket{\ell, \mathfrak{m}'} \,.
 }

We can then consider the basis of functions on $\grSU(2)$ to be
 \es{BasisStates}{
  \Psi_{\ell, \mathfrak{m}_L, \mathfrak{m}_R}(U) = \sqrt{\frac{2 \ell + 1}{2 \pi^2 }} \langle \ell, \mathfrak{m}_L | U | \ell, \mathfrak{m}_R \rangle \,.
 }
It can be checked that these functions are orthonormal with respect to the Haar measure $d\mu(U)$ on $\grSU(2)$:
 \es{Ortho}{
  \int d\mu(U) \, \Psi_{\ell', \mathfrak{m}_L', \mathfrak{m}_R'}(U)^*  \Psi_{\ell, \mathfrak{m}_L, \mathfrak{m}_R}(U)  = \delta_{\ell \ell'} \delta_{\mathfrak{m}_L \mathfrak{m}_L'} \delta_{\mathfrak{m}_R \mathfrak{m}_R'} \,.
 }
Starting with $\pi_\mu \Psi_{\ell, \mathfrak{m}_L, \mathfrak{m}_R}(U) = \Psi_{\ell, \mathfrak{m}_L, \mathfrak{m}_R}(i \partial_\mu U)$, we can show that the angular momentum generators act as
 \es{GenActionOnStates}{
  L^A \Psi_{\ell, \mathfrak{m}_L, \mathfrak{m}_R}(U) = \Psi_{\ell, \mathfrak{m}_L, \mathfrak{m}_R}(T^A U) \,, \qquad
   R^A \Psi_{\ell, \mathfrak{m}_L, \mathfrak{m}_R}(U) = \Psi_{\ell, \mathfrak{m}_L, \mathfrak{m}_R}( U T^A) \,.
 }
Given \eqref{TAction}, we then have
 \es{GenAction2}{
  L^A \Psi_{\ell, \mathfrak{m}_L, \mathfrak{m}_R} =  T^A_{\mathfrak{m}_L \mathfrak{m}_L'} \Psi_{\ell, \mathfrak{m}_L', \mathfrak{m}_R} \,, \qquad
  R^A \Psi_{\ell, \mathfrak{m}_L, \mathfrak{m}_R} =  T^A_{\mathfrak{m}_R' \mathfrak{m}_R} \Psi_{\ell, \mathfrak{m}_L, \mathfrak{m}_R'}  \,.
 }
In particular, this relation with $A=3$ shows that $\Psi_{\ell, \mathfrak{m}_L, \mathfrak{m}_R}$ are eigenstates of $L^3$ and $R^3$ with eigenvalues $\mathfrak{m}_L$ and $\mathfrak{m}_R$, respectively.  Note that the relations \eqref{GenAction2} are consistent with the commutation relations in the first line of \eqref{LRUCommut}.

\section{Decomposition of the spinor representation}\label{app:rep_decomp}

In our lattice model for the $\grSU(N_c)$ gauge theory coupled to Majorana adjoint fermions on $N$ sites, the Hilbert space of the fermions before imposing gauge invariance transforms in the spinor representation of $\so(N(N_c^2-1))$. This representation is reducible into the sum of two half-spinor representations, which corresponds to the decomposition of the Hilbert space into bosonic and fermionic states, as explained in Section \ref{GAUGEINVSTATES}. Here we will ignore the reducibility of the spinor representation of $\so(M)$ when $M$ is even, and denote this representation by $\spin(M)$. We denote the defining vector representation of $\so(M)$ by $\vector(M)$, and the adjoint representation of $\su(M)$ by $\adj(M)$.

To construct gauge-invariant states, we need to understand how $\spin(N(N_c^2 - 1))$ branches under
\begin{equation}
	\su(N_c)^N \hookrightarrow \so(N_c^2 - 1)^N \hookrightarrow \so(N(N_c^2 - 1)) \,.
\end{equation}
The embedding is fixed by requiring that $\vector(N_c^2 - 1)$ branches into $\adj(N_c)$, and that
\begin{equation}\label{eq:branching_requirement}
	\small \vector(N(N_c^2 - 1)) \mapsto (\vector(N_c^2 - 1), 1, \ldots, 1) \oplus (1, \vector(N_c^2 - 1), \ldots, 1) \oplus \cdots \oplus (1, 1, \ldots, \vector(N_c^2 - 1)) \,.
\end{equation}

We can start by understanding how $\spin(N(N_c^2 - 1))$ branches under the embedding $\so(N_c^2 - 1)^N \hookrightarrow \so(N(N_c^2 - 1))$. In an orthogonal basis, the weights of $\vector(M)$ are
\begin{equation}
	\Big(\underbrace{0, 0, \ldots, \pm 1, \ldots, 0}_{\lfloor M/2\rfloor}\Big) \,,
\end{equation}
and additionally $(0,\ldots,0)$ if $M$ is odd. The condition \eqref{eq:branching_requirement} tells us how to map the $\frac{N(N_c^2 -1)}{2}$ weights of $\so(N(N_c^2 -1))$ into the $N\lfloor \frac{N_c^2 - 1}{2}\rfloor$ weights of $\so(N_c^2 - 1)^N$. The projection matrix takes the form
\begin{equation}\label{eq:proj1}
	P_1 = \begin{pmatrix} \mathbbm{1}_{N\lfloor (N_c^2 - 1)/2\rfloor} & 0_{N\lfloor (N_c^2 - 1)/2\rfloor \times N/2} \end{pmatrix}\quad (N_c\text{ even}), \qquad P_1 = \mathbbm{1}_{N(N_c^2 - 1)/2} \quad (N_c\text{ odd}) \,,
\end{equation}
where $\mathbbm{1}_d$ is the $d\times d$ identity matrix and $0_{m\times n}$ is the $m\times n$ zero matrix.

In the same basis, the weights of $\spin(M)$ are
\begin{equation}
	\Big(\underbrace{\pm \frac{1}{2}, \pm \frac{1}{2}, \ldots, \pm\frac{1}{2}}_{\lfloor M/2\rfloor}\Big) \,.
\end{equation}
Using \eqref{eq:proj1}, we find
 \es{SpinNDecomp}{
	\spin(N(N_c^2 - 1)) \mapsto \left(\begin{cases} 2^{N/2} & N_c\text{ even} \\ 1 & N_c\text{ odd} \end{cases}\right)\cdot \spin(N_c^2 - 1)^N \,.
 }

Now we can work out the branching of $\spin(N_c^2 - 1)$ under the embedding $\su(N_c)\hookrightarrow \so(N_c^2 - 1)$. Since the vector of $\so(N_c^2 - 1)$ has to branch into $\adj(N_c)$, the projection matrix can be taken as
 \es{P2}{
	P_2 = \begin{pmatrix} \vec{\alpha}_1 & \vec{\alpha}_2 & \cdots & \vec{\alpha}_{N_c(N_c - 1)/2} & 0_{(N_c - 1) \times \lfloor (N_c - 1)/2\rfloor } \end{pmatrix} \,,
 }
where the $\vec{\alpha}_i$ are positive roots of $\su(N_c)$ in the orthogonal basis. Using this projection, we find that the weights of $\spin(N_c^2 - 1)$ map to
 \es{WeightsR}{
	\left\lbrace \vec{\rho}, \vec{\rho} - \vec{\alpha}_1, \vec{\rho} - \vec{\alpha}_2, \ldots, \vec{\rho} - \sum_i \vec{\alpha}_i\right\rbrace \,,
 }
where $\vec{\rho} = \frac{1}{2}\sum_i \vec{\alpha}_i$ is the Weyl vector of $\su(N_c)$, each with multiplicity $2^{\lfloor (N_c - 1)/2\rfloor}$ coming from the zero columns in \eqref{P2}.  Not counting this multiplicity, the set of weights in \eqref{WeightsR} are precisely those of the representation ${\bf R}$ defined in \eqref{Diagram} whose highest weight is $\vec{\rho}$.  (One can check that this representation has dimension $2^{N_c (N_c-1)/2}$ and there are precisely this many weights in \eqref{WeightsR}.  Thus, under the embedding $\su(N_c)\hookrightarrow \so(N_c^2 - 1)$, we have
 \es{DecompSpin}{
	\spin(N_c^2 - 1) \mapsto 2^{\lfloor (N_c - 1)/2\rfloor} {\bf R} \,.
 }

Combining \eqref{SpinNDecomp} with \eqref{DecompSpin}, we find
\begin{equation}
	\spin(N(N_c^2 - 1)) \mapsto 2^{N(N_c - 1)/2} \left({\bf R}, {\bf R}, \ldots, {\bf R}\right) \,,
\end{equation}
as in \eqref{Decomp}.

\section{The action of ${\cal O}_n$ on gauge-invariant states}
\label{GAUGEINVAPPENDIX}

In this Appendix we explain the derivation of \eqref{OnAction}, namely how the operator ${\cal O}_n \equiv S_n^A U_n^{AB} S_{n+1}^B$ defined in \eqref{eq:even_pauli} acts on the gauge-invariant states \eqref{GaugeInv}.  Since ${\cal O}_n$ only involves the states on sites $n$ and $n+1$, let us strip out the factors from \eqref{GaugeInv} that do not involve $m_n$ or $m_{n+1}$, and define
 \es{ReducedStateClutter}{
  \psi^{\ell_{(n-1)R},\ell_n, \ell_{(n+1)L}}_{\mathfrak{m}_{(n-1)R}, \mathfrak{m}_{(n+1)L} } \equiv \sum_{\substack{m_n, m_{n+1},\\ \mathfrak{m}_{nL}, \mathfrak{m}_{nR}}}   \ket{\half, m_n}  \ket{\half, m_{n+1}}     
  C^{\ell_{n-1} \frac{1}{2} \ell_n}_{\mathfrak{m}_{(n-1)R} m_n \mathfrak{m}_{nL}}
  C^{\ell_n \frac{1}{2} \ell_{n+1}}_{\mathfrak{m}_{nR} m_{n+1} \mathfrak{m}_{(n+1)L}} \frac{\ket{\ell_n, \mathfrak{m}_{nL}, \mathfrak{m}_{nR} }}{\sqrt{2 \ell_n + 1}}  \,.
}
In other words, we only impose the Gauss law on sites $n$ and $n+1$ and consider the quantum numbers $\mathfrak{m}_{(n+1)R}$ and $\mathfrak{m}_{(n-1)L}$ as fixed.  The action of ${\cal O}_n$ on \eqref{GaugeInv} can be straightforwardly inferred from its action on this state.

In order to declutter \eqref{ReducedStateClutter}, let us use the simplified notation
 \es{SimpNotation}{
 & m_n \to m_1 \,, \qquad m_{n+1} \to m_2 \\
  &(\ell_{n-1}, \mathfrak{m}_{(n-1)R}) \to (\ell_l, \mathfrak{m}_l) \,, \\
  &(\ell_{n}, \mathfrak{m}_{nL}, \mathfrak{m}_{nR}) \to (\ell, \mathfrak{m}_L, \mathfrak{m}_R) \,, \\
  &(\ell_{n+1}, \mathfrak{m}_{(n+1)L}) \to (\ell_r, \mathfrak{m}_r) \,, \\
  &U_n \to U \,, \qquad {\cal O}_n \to {\cal O} = S_1^A U^{AB} S_2^B \,.
 } 
Thus, we consider the state
 \es{ReducedState}{
  \psi^{\ell_l, \ell, \ell_r}_{\mathfrak{m}_l, \mathfrak{m}_r} \equiv \sum_{\substack{m_1, m_2,\\ \mathfrak{m}_{L}, \mathfrak{m}_{R}}}   \ket{\half, m_1}  \ket{\half, m_2}     
  C^{\ell_l \frac{1}{2} \ell}_{\mathfrak{m}_{l} m_1 \mathfrak{m}_{L}}
  C^{\ell \frac{1}{2} \ell_r}_{\mathfrak{m}_{R} m_2 \mathfrak{m}_{r}} \frac{\ket{\ell, \mathfrak{m}_{L}, \mathfrak{m}_{R} }}{\sqrt{2 \ell + 1}} \,.
} 

Writing the Clebsch-Gordan coefficients in \eqref{ReducedState} as matrix elements, passing to the position representation for the bosonic state, and defining $\psi^{\ell_l, \ell, \ell_r}_{\mathfrak{m}_r} \equiv \sum_{\mathfrak{m}_l} \psi^{\ell_l, \ell, \ell_r}_{\mathfrak{m}_l, \mathfrak{m}_r} \ket{\ell_l, \mathfrak{m}_l}$, we have
 \es{ReducedState2}{
  \psi^{\ell_l, \ell, \ell_r}_{\mathfrak{m}_r}(U) 
    &\equiv 
      \frac{1}{\sqrt{2 \pi^2}} \sum_{\substack{m_1, m_2,\\ \mathfrak{m}_l, \mathfrak{m}_{L}, \mathfrak{m}_{R}}}
      \ket{\ell_l, \mathfrak{m}_l} \ket{\half, m_1} 
      \bkt{\ell_l, \mathfrak{m}_l, \half, m_1}{\ell, \mathfrak{m}_{L}}  \\
       &\qquad \qquad \qquad\qquad\times \bkt{\ell, \mathfrak{m}_{L}}{U | \ell, \mathfrak{m}_R}  
          \ket{\half, m_2} \bkt{\ell, \mathfrak{m}_R, \half, m_2}{\ell_r, \mathfrak{m}_r}   \,.
  }
To simplify the following analysis, we can pass to a new basis $\ket{\ell, \mathfrak{m}_R} \to U^{-1} \ket{\ell, \mathfrak{m}_R}$, $\ket{\half, m_2} \to U^{-1} \ket{\half, m_2}$, $\ket{\ell_r, \mathfrak{m}_r} \to U^{-1} \ket{\ell_r, \mathfrak{m}_r}$, where $U^{-1}$ acts in the appropriate $\grSU(2)$ representation.  After this change of basis, we have $\bkt{\ell, \mathfrak{m}_{L}}{ \mathds{1} | \ell, \mathfrak{m}_R} = \delta_{\mathfrak{m}_L \mathfrak{m}_R}$, so (denoting $\mathfrak{m}_L = \mathfrak{m}_R = \mathfrak{m}$)
 \es{ReducedState3}{
  \psi^{\ell_l, \ell, \ell_r}_{\mathfrak{m}_r} 
    \equiv 
      \frac{1}{\sqrt{2 \pi^2}} \sum_{\substack{m_1, m_2,\\ \mathfrak{m}_l,  \mathfrak{m}}}
      \ket{\ell_l, \mathfrak{m}_l}  \ket{\half, m_1}   \bkt{\ell_l, \mathfrak{m}_l, \half, m_1}{\ell, \mathfrak{m}}
      \ket{\half, m_2}  \bkt{\ell, \mathfrak{m}, \half, m_2}{\ell_r, \mathfrak{m}_r} \,.       
  }
The operator ${\cal O}$ in \eqref{SimpNotation} whose action we want to determine simplifies to ${\cal O} = S_1^A S_2^A$.

The equation \eqref{ReducedState3} has the manifest structure of addition of three angular momenta, for which $(\ell_l \otimes \half) \otimes \half \to \ell \otimes \half \to \ell_r$.  In other words, denoting by $\vec{J}_l$, $\vec{J}_1$, $\vec{J}_2$, $\vec{J}$, $\vec{J}_r$ the angular momentum operators acting on the states with magnetic quantum numbers $m_l$, $m_1$, $m_2$, $m$, and $m_r$, respectively, we have
 \es{JAddition}{
  \vec{J} = \vec{J}_l + \vec{J}_1 \,, \qquad \vec{J}_r = \vec{J} + \vec{J}_2 \,.
 } 
In standard notation, \eqref{ReducedState3} can also be written as
 \es{ReducedState4}{
  \psi^{\ell_l, \ell, \ell_r}_{\mathfrak{m}_r} 
    \equiv 
      \frac{1}{\sqrt{2 \pi^2}} \ket{\left((\ell_l \half) \ell \half\right) \ell_r \mathfrak{m}_r}      
  }
and we would like to determine the action of the operator ${\cal O} = \vec{J}_1 \cdot \vec{J}_2$.  Since ${\cal O } = \frac{1}{2} (J_{12}^2 - J_1^2 - J_2^2)$ where $\vec{J}_{12} = \vec{J}_1 + \vec{J}_2$, there is another way of writing the state $\psi_{\mathfrak{m}_r} $ in which the action of ${\cal O}$ is trivial, and where we first multiply together the spin-$1/2$ states into states of angular momentum $s$:  $\ell_l \otimes (\half \otimes \half) \to \ell_l \otimes s \to \ell_r$:
\es{Anotherway}{
  \psi^{\ell_l, \ell, \ell_r}_{\mathfrak{m}_r} 
    = 
      \frac{1}{\sqrt{2 \pi^2}} \sum_{s = 0, 1} 
       \sqrt{(2s  + 1)(2\ell+1)} W(\ell_l \half \ell_r \half; \ell s) \ket{\left(\ell_l, (\half \half) s \right) \ell_r \mathfrak{m}_r}      
  }
where $W$ are Racah coefficients.  Acting with ${\cal O}$ on each term then gives a factor of $\frac{1}{2} \left(s(s+1) - \frac 32\right) $, so in total we have 
 \es{OAction}{
  {\cal O}\psi^{\ell_l, \ell, \ell_r}_{\mathfrak{m}_r} = \sum_{\ell'}  f(\ell_l, \ell_f; \ell', \ell) \psi^{\ell_l, \ell', \ell_r}_{\mathfrak{m}_r} \,,
 }
with 
\es{fExplicit}{
  f(\ell_l, \ell_r; \ell', \ell)  = \sum_{s=0,1} (2s+1) \sqrt{(2 \ell + 1)(2 \ell' + 1)} \frac 12 \left[s(s+1) - \frac 32 \right] W(\ell_l \half \ell_r \half; \ell s) W(\ell_l \half \ell_r \half; \ell' s)  \,.
 }
Note that for $s=0, 1$, we have $\frac 12 (2s+1) \left[s(s+1) - \frac 32 \right] = (-1)^{s + 1} \frac 34$.  In terms of 6$j$-symbols, the equation above can also be written as
 \es{fExpAgain}{
  f(\ell_l, \ell_r; \ell', \ell)  = \sum_{s=0,1} \frac 34  (-1)^{2\ell_l +2 \ell_r + s + 1} \sqrt{(2 \ell + 1)(2 \ell' + 1)} 
   \begin{Bmatrix} \ell_l & \half & \ell \\
   \half & \ell_r & s 
   \end{Bmatrix}  \begin{Bmatrix} \ell_l & \half & \ell' \\
   \half & \ell_r & s 
   \end{Bmatrix}   \,.
  }

Note that $2 \ell_l$ must have the same parity as $2 \ell_r$, so $(-1)^{2 \ell_l + 2 \ell_r} = 1$.  Moreover, we can insert inside the sum in \eqref{fExpAgain} the identity
 \es{id2sp1}{
  1 = 2 (2s + 1) \begin{Bmatrix} \frac 12 & \frac 12 & s \\
  \frac 12 & \frac 12 & 1 \end{Bmatrix} \,.
 }
Then, using an identity that relates a sum over products of three 6$j$-symbols to a single product of two 6$j$-symbols, we can show that
 \es{Gotf}{
 f(\ell_l, \ell_r; \ell', \ell)  = (-1)^{\ell_l + \ell + \ell' + \ell_r} 
  \frac 32 \sqrt{(2 \ell + 1) (2 \ell' + 1)} 
   \begin{Bmatrix}
    \frac 12 & \frac 12 & 1 \\
    \ell' & \ell & \ell_l 
    \end{Bmatrix}
     \begin{Bmatrix}
    \frac 12 & \frac 12 & 1 \\
    \ell' & \ell & \ell_r
    \end{Bmatrix} \,.
 } 
This is the expression for $f$ that should be substituted in \eqref{OnAction} in the main text.

\section{Large mass limit on the lattice}\label{app:large_mass}

We will now explain how the large mass limit $m\gg \gym$ considered in section \ref{sec:large_mass} can also be understood analytically in the lattice model, if we simultaneously approach the continuum and heavy fermion limit as $a^{-1} = |m| \gg \gym$ with $L=aN$ fixed. Even for finite $a$, setting $a^{-1} = |m|$ offers a significant simplification in the lattice Hamiltonian \eqref{eq:lattice_hamiltonian}, because either the odd or even hopping terms cancel depending on the sign of~$m$.

We can begin with a positive mass $m=a^{-1}$. The Hamiltonian reduces to
\begin{align}
	H = a^{-1}\left[- i\sum_{k=0}^{N/2-1} \chi_{2k}^AU^{AB}_{2k}\chi_{2n+1}^B+\frac{(\gym a)^2}{2}\sum_{n=0}^N L^A_nL^A_n \right] \equiv a^{-1}[ W_\text{even}+(\gym a)^2W_\text{gauge}] \,.
\end{align}
In the limit $a\to 0$, we can treat the dimensionless gauge kinetic term $W_\text{gauge}$ as a perturbation of the sum of the fermion kinetic and mass terms, denoted $W_\text{even}$. 

Due to the cancellation between the fermion kinetic and mass terms for $m = a^{-1}$, the degenerate ground states of $W_\text{even}$ can be determined exactly. Writing out $W_\text{even}$ using the basis \eqref{eq:majoranas}, we find $\frac{N}{2}$ terms
\begin{align}
	W_\text{even} = 2\sum_{k=0}^{N/2-1}Z_k\otimes S^A_{2k}U^{AB}_{2k}S^B_{2k+1} \,.
\end{align}
As detailed in Appendix \ref{GAUGEINVAPPENDIX}, we can effectively set $U^{AB}_{2k}=\delta^{AB}$ which leaves a simple Hamiltonian of decoupled terms $2Z_k\otimes S^A_{2k}S^A_{2k+1}$ on every even link. The factor $S^A_{2k}S^A_{2k+1}$ has eigenvalues $-\frac{3}{4}$ or $\frac{1}{4}$ depending on whether the spin-$\frac{1}{2}$s are put in a singlet or triplet configuration. Thus, the ground states will have the $\grSU(2)$ spins arranged pairwise in singlets, and all qubits in the $s_k=+1$ state. 

This ground state space is infinitely degenerate because we also have to include the representations on the links. If we fix the representation on one of the odd links to $\ell\in \mathbb{Z}/2$, then the representations on all odd links are $\ell$ due to the Gauss law, and  hence the representations on even links are $\ell \pm \frac{1}{2}$. To find the explicit form of the state, we take the $2\times 2$ Hamiltonian corresponding to the action of $S^A_{2k}U^{AB}_{2k}S^B_{2k+1}$ on an even link in the basis $\{\ket{\ldots,\ell,\ell+\frac{1}{2},\ell,\ldots},\ket{\ldots,\ell,\ell-\frac{1}{2},\ell,\ldots}\}$:
\begin{align}
	S^A_{2k}U^{AB}_{2k}S^B_{2k+1} = \begin{pmatrix}
		-\frac{2\ell+3}{4(2\ell +1)}&\frac{\sqrt{\ell(\ell+1)}}{2\ell+1}\\
		\frac{\sqrt{\ell(\ell+1)}}{2\ell+1} & -\frac{2\ell-1}{4(2\ell+1)}
	\end{pmatrix} \,, 
\end{align}
which follows from the matrix elements derived in Appendix \ref{GAUGEINVAPPENDIX}. The eigenvector with eigenvalue $-\frac{3}{4}$ is
\begin{align}
	S^A_{2k}S^A_{2k+1}\begin{pmatrix}
		c_{\ell,+}\\
		c_{\ell,-}
	\end{pmatrix}=-\frac{3}{4}\begin{pmatrix}
	c_{\ell,+}\\
	c_{\ell,-}
	\end{pmatrix}\,,\qquad c_{\ell,+} = -\sqrt{\frac{1+\ell}{1+2\ell}}\,,\qquad c_{\ell,-} = \sqrt{\frac{\ell}{1+\ell}} \,.
\end{align}
Using these coefficients, we can write down the full set of degenerate ground states parameterized by the representation $\ell$ on the odd link as 
\begin{align}\label{eq:even_state}
	\ket{\ell} = \sum_{m_n=\pm\frac{1}{2}}c_{\ell,+}^{\frac{N}{4}+\sum_n m_n}c^{\frac{N}{4}-\sum_n m_n}_{\ell,-}\ket{\ell + m_0,\ell,\ldots,\ell, \ell+m_{N/2-1}}\otimes \ket{1,\ldots,1} \,.
\end{align}
These states all have unperturbed energies of
\begin{equation}
	W_\text{even} \ket{\ell} = -\frac{3N}{4} \ket{\ell} \,.
\end{equation}
The matrix elements of the perturbing term are simply
\begin{align}\label{eq:large_mass_splitting}
	\braket{\ell|W_\text{gauge}|\ell'} = \delta_{\ell,\ell'}\frac{N}{2}\left[\ell(\ell+1)+\frac{3}{8}\right] \,.
\end{align}
from which we can read off the perturbed spectrum. After restoring the factor of $\gym^2 a$, this gives the spectrum \eqref{eq:pure_su2} of the pure SU(2) theory. Since the ground states are parametrized by odd link representations $\ell\in\mathbb{Z}$, this gives the gaps \eqref{eq:trivial_gaps}.

If we instead take $m = -a^{-1}$, then we have a very similar problem, except the singlet pairs are on sites $(2n-1, 2n)$ and the degenerate ground state space is parametrized by a representation $\ell \in \mathbb{Z} + \frac{1}{2}$ on an even link. We clearly recover the same splitting \eqref{eq:large_mass_splitting}, except with $\ell \in \mathbb{Z} + \frac{1}{2}$, which leads to \eqref{eq:nontrivial_gaps}. 

\bibliographystyle{ssg}
\bibliography{Adjointmain}

\begingroup\raggedright\begin{thebibliography}{10}

\bibitem{Wilson:1974sk}
K.~G. Wilson, ``{Confinement of Quarks},'' {\em Phys. Rev. D} {\bf 10} (1974)
  2445--2459.

\bibitem{Schwinger:1962tp}
J.~S. Schwinger, ``{Gauge Invariance and Mass. 2.},'' {\em Phys. Rev.} {\bf
  128} (1962) 2425--2429.

\bibitem{Dalley:1992yy}
S.~Dalley and I.~R. Klebanov, ``{String spectrum of (1+1)-dimensional large N
  QCD with adjoint matter},'' {\em Phys. Rev. D} {\bf 47} (1993) 2517--2527,
  \href{https://arxiv.org/abs/hep-th/9209049}{{\tt hep-th/9209049}}.

\bibitem{Kutasov:1993gq}
D.~Kutasov, ``{Two-dimensional QCD coupled to adjoint matter and string
  theory},'' {\em Nucl. Phys. B} {\bf 414} (1994) 33--52,
  \href{https://arxiv.org/abs/hep-th/9306013}{{\tt hep-th/9306013}}.

\bibitem{Bhanot:1993xp}
G.~Bhanot, K.~Demeterfi, and I.~R. Klebanov, ``{(1+1)-dimensional large N QCD
  coupled to adjoint fermions},'' {\em Phys. Rev. D} {\bf 48} (1993)
  4980--4990, \href{https://arxiv.org/abs/hep-th/9307111}{{\tt
  hep-th/9307111}}.

\bibitem{Delmastro:2021otj}
D.~Delmastro, J.~Gomis, and M.~Yu, ``{Infrared phases of 2d QCD},'' {\em JHEP}
  {\bf 02} (2023) 157, \href{https://arxiv.org/abs/2108.02202}{{\tt
  2108.02202}}.

\bibitem{Gross:1995bp}
D.~J. Gross, I.~R. Klebanov, A.~V. Matytsin, and A.~V. Smilga, ``{Screening
  versus confinement in (1+1)-dimensions},'' {\em Nucl. Phys. B} {\bf 461}
  (1996) 109--130, \href{https://arxiv.org/abs/hep-th/9511104}{{\tt
  hep-th/9511104}}.

\bibitem{Gross:1997mx}
D.~J. Gross, A.~Hashimoto, and I.~R. Klebanov, ``{The Spectrum of a large N
  gauge theory near transition from confinement to screening},'' {\em Phys.
  Rev. D} {\bf 57} (1998) 6420--6428,
  \href{https://arxiv.org/abs/hep-th/9710240}{{\tt hep-th/9710240}}.

\bibitem{Komargodski:2020mxz}
Z.~Komargodski, K.~Ohmori, K.~Roumpedakis, and S.~Seifnashri, ``{Symmetries and
  strings of adjoint QCD$_{2}$},'' {\em JHEP} {\bf 03} (2021) 103,
  \href{https://arxiv.org/abs/2008.07567}{{\tt 2008.07567}}.

\bibitem{Dempsey:2021xpf}
R.~Dempsey, I.~R. Klebanov, and S.~S. Pufu, ``{Exact symmetries and threshold
  states in two-dimensional models for QCD},'' {\em JHEP} {\bf 10} (2021) 096,
  \href{https://arxiv.org/abs/2101.05432}{{\tt 2101.05432}}.

\bibitem{Witten:1978ka}
E.~Witten, ``{$\theta$ Vacua in Two-dimensional Quantum Chromodynamics},'' {\em
  Nuovo Cim. A} {\bf 51} (1979) 325.

\bibitem{Smilga:1994hc}
A.~V. Smilga, ``{Instantons and fermion condensate in adjoint QCD in
  two-dimensions},'' {\em Phys. Rev. D} {\bf 49} (1994) 6836--6848,
  \href{https://arxiv.org/abs/hep-th/9402066}{{\tt hep-th/9402066}}.

\bibitem{Lenz:1994du}
F.~Lenz, M.~A. Shifman, and M.~Thies, ``{Quantum mechanics of the vacuum state
  in two-dimensional QCD with adjoint fermions},'' {\em Phys. Rev. D} {\bf 51}
  (1995) 7060--7082, \href{https://arxiv.org/abs/hep-th/9412113}{{\tt
  hep-th/9412113}}.

\bibitem{Cherman:2019hbq}
A.~Cherman, T.~Jacobson, Y.~Tanizaki, and M.~\"Unsal, ``{Anomalies, a mod 2
  index, and dynamics of 2d adjoint QCD},'' {\em SciPost Phys.} {\bf 8} (2020),
  no.~5 072, \href{https://arxiv.org/abs/1908.09858}{{\tt 1908.09858}}.

\bibitem{tHooft:1974pnl}
G.~'t~Hooft, ``A {Two}-{Dimensional} {Model} for {Mesons},'' {\em Nucl. Phys.
  B} {\bf 75} (1974) 461--470.

\bibitem{Dempsey:2022uie}
R.~Dempsey, I.~R. Klebanov, L.~L. Lin, and S.~S. Pufu, ``{Adjoint Majorana
  QCD$_{2}$ at finite N},'' {\em JHEP} {\bf 04} (2023) 107,
  \href{https://arxiv.org/abs/2210.10895}{{\tt 2210.10895}}.

\bibitem{Trittmann:2023dar}
U.~Trittmann, ``{Solving two-dimensional adjoint QCD with a basis-function
  approach},'' \href{https://arxiv.org/abs/2307.15212}{{\tt 2307.15212}}.

\bibitem{Kogut:1974ag}
J.~B. Kogut and L.~Susskind, ``{Hamiltonian Formulation of Wilson's Lattice
  Gauge Theories},'' {\em Phys. Rev. D} {\bf 11} (1975) 395--408.

\bibitem{Kitaev:2000nmw}
A.~Kitaev, ``{Unpaired Majorana fermions in quantum wires},'' {\em Phys. Usp.}
  {\bf 44} (2001), no.~10S 131--136,
  \href{https://arxiv.org/abs/cond-mat/0010440}{{\tt cond-mat/0010440}}.

\bibitem{Seiberg:2023cdc}
N.~Seiberg and S.-H. Shao, ``{Majorana chain and Ising model --
  (non-invertible) translations, anomalies, and emanant symmetries},''
  \href{https://arxiv.org/abs/2307.02534}{{\tt 2307.02534}}.

\bibitem{Dempsey:2022nys}
R.~Dempsey, I.~R. Klebanov, S.~S. Pufu, and B.~Zan, ``{Discrete chiral symmetry
  and mass shift in the lattice Hamiltonian approach to the Schwinger model},''
  {\em Phys. Rev. Res.} {\bf 4} (2022), no.~4 043133,
  \href{https://arxiv.org/abs/2206.05308}{{\tt 2206.05308}}.

\bibitem{Dempsey:2023gib}
R.~Dempsey, I.~R. Klebanov, S.~S. Pufu, B.~T. S\o{}gaard, and B.~Zan, ``{Phase
  Diagram of the Two-Flavor Schwinger Model at Zero Temperature},'' {\em Phys.
  Rev. Lett.} {\bf 132} (2024), no.~3 031603,
  \href{https://arxiv.org/abs/2305.04437}{{\tt 2305.04437}}.

\bibitem{Hamer:1981yq}
C.~J. Hamer, ``{SU(2) {Yang-Mills} Theory in (1+1)-dimensions: A Finite Lattice
  Approach},'' {\em Nucl. Phys. B} {\bf 195} (1982) 503--521.

\bibitem{Banuls:2017ena}
M.~C. Ba\~nuls, K.~Cichy, J.~I. Cirac, K.~Jansen, and S.~K\"uhn, ``{Efficient
  basis formulation for 1+1 dimensional SU(2) lattice gauge theory: Spectral
  calculations with matrix product states},'' {\em Phys. Rev. X} {\bf 7}
  (2017), no.~4 041046, \href{https://arxiv.org/abs/1707.06434}{{\tt
  1707.06434}}.

\bibitem{Pauli:1985ps}
H.~C. Pauli and S.~J. Brodsky, ``Discretized {Light} {Cone} {Quantization}:
  {Solution} to a {Field} {Theory} in {One} {Space} {One} {Time}
  {Dimensions},'' {\em Phys. Rev. D} {\bf 32} (1985) 2001.

\bibitem{Demeterfi:1993rs}
K.~Demeterfi, I.~R. Klebanov, and G.~Bhanot, ``Glueball spectrum in a
  (1+1)-dimensional model for {QCD},'' {\em Nuclear Physics B} {\bf 418} (Apr.,
  1994) 15--29. arXiv: hep-th/9311015.

\bibitem{Boorstein:1993nd}
J.~Boorstein and D.~Kutasov, ``Symmetries and {Mass} {Splittings} in
  {QCD}\$\_2\$ {Coupled} to {Adjoint} {Fermions},'' {\em Nuclear Physics B}
  {\bf 421} (June, 1994) 263--277. arXiv:hep-th/9401044.

\bibitem{Kutasov:1994xq}
D.~Kutasov and A.~Schwimmer, ``{Universality in two-dimensional gauge
  theory},'' {\em Nucl. Phys. B} {\bf 442} (1995) 447--460,
  \href{https://arxiv.org/abs/hep-th/9501024}{{\tt hep-th/9501024}}.

\bibitem{Katz:2013qua}
E.~Katz, G.~Marques~Tavares, and Y.~Xu, ``{Solving 2D QCD with an adjoint
  fermion analytically},'' {\em JHEP} {\bf 05} (2014) 143,
  \href{https://arxiv.org/abs/1308.4980}{{\tt 1308.4980}}.

\bibitem{Popov:2022vud}
F.~K. Popov, ``Supersymmetry in {QCD}\$\_2\$ coupled to fermions,'' {\em
  Physical Review D} {\bf 105} (Apr., 2022) 074005. arXiv:2202.04017 [hep-th].

\bibitem{Antonuccio:1998uz}
F.~Antonuccio and S.~Pinsky, ``On the {Transition} from {Confinement} to
  {Screening} in {QCD}\_\{1+1\} {Coupled} to {Adjoint} {Fermions} at {Finite}
  {N},'' {\em Physics Letters B} {\bf 439} (Oct., 1998) 142--149. arXiv:
  hep-th/9805188.

\bibitem{Cheng:2022sgb}
M.~Cheng and N.~Seiberg, ``{Lieb-Schultz-Mattis, Luttinger, and 't Hooft -
  anomaly matching in lattice systems},'' {\em SciPost Phys.} {\bf 15} (2023),
  no.~2 051, \href{https://arxiv.org/abs/2211.12543}{{\tt 2211.12543}}.

\bibitem{Antonuccio:1998zp}
F.~Antonuccio, O.~Lunin, and S.~Pinsky, ``{On exact supersymmetry in DLCQ},''
  {\em Phys. Lett. B} {\bf 442} (1998) 173--179,
  \href{https://arxiv.org/abs/hep-th/9809165}{{\tt hep-th/9809165}}.

\bibitem{Dubovsky:2018dlk}
S.~Dubovsky, ``{A Simple Worldsheet Black Hole},'' {\em JHEP} {\bf 07} (2018)
  011, \href{https://arxiv.org/abs/1803.00577}{{\tt 1803.00577}}.

\bibitem{Banks:1975gq}
T.~Banks, L.~Susskind, and J.~B. Kogut, ``{Strong Coupling Calculations of
  Lattice Gauge Theories: (1+1)-Dimensional Exercises},'' {\em Phys. Rev. D}
  {\bf 13} (1976) 1043.

\bibitem{Hamer:1997dx}
C.~J. Hamer, W.-h. Zheng, and J.~Oitmaa, ``{Series expansions for the massive
  Schwinger model in Hamiltonian lattice theory},'' {\em Phys. Rev. D} {\bf 56}
  (1997) 55--67, \href{https://arxiv.org/abs/hep-lat/9701015}{{\tt
  hep-lat/9701015}}.

\bibitem{Hamer:1982mx}
C.~J. Hamer, J.~B. Kogut, D.~P. Crewther, and M.~M. Mazzolini, ``{The Massive
  Schwinger Model on a Lattice: Background Field, Chiral Symmetry and the
  String Tension},'' {\em Nucl. Phys. B} {\bf 208} (1982) 413--438.

\bibitem{messiah1999quantum}
A.~Messiah, {\em Quantum Mechanics}.
\newblock Dover books on physics. Dover Publications, 1999.

\bibitem{petsc-user-ref}
S.~Balay, S.~Abhyankar, M.~F. Adams, J.~Brown, P.~Brune, K.~Buschelman,
  L.~Dalcin, V.~Eijkhout, W.~D. Gropp, D.~Karpeyev, D.~Kaushik, M.~G. Knepley,
  D.~A. May, L.~C. McInnes, R.~T. Mills, T.~Munson, K.~Rupp, P.~Sanan, B.~F.
  Smith, S.~Zampini, H.~Zhang, and H.~Zhang, ``{PETS}c Users Manual,'' Tech.
  Rep. ANL-95/11 - Revision 3.11, Argonne National Laboratory, 2019.

\bibitem{petsc-efficient}
S.~Balay, W.~D. Gropp, L.~C. McInnes, and B.~F. Smith, ``Efficient Management
  of Parallelism in Object Oriented Numerical Software Libraries,'' in {\em
  Modern Software Tools in Scientific Computing} (E.~Arge, A.~M. Bruaset, and
  H.~P. Langtangen, eds.), pp.~163--202, Birkh{\"{a}}user Press, 1997.

\bibitem{slepc-toms}
V.~Hernandez, J.~E. Roman, and V.~Vidal, ``{SLEPc}: A Scalable and Flexible
  Toolkit for the Solution of Eigenvalue Problems,'' {\em {ACM} Trans. Math.
  Software} {\bf 31} (2005), no.~3 351--362.

\bibitem{slepc-manual}
J.~E. Roman, C.~Campos, L.~Dalcin, E.~Romero, and A.~Tomas, ``{SLEPc} Users
  Manual,'' Tech. Rep. DSIC-II/24/02 - Revision 3.16, D. Sistemes Inform\`atics
  i Computaci\'o, Universitat Polit\`ecnica de Val\`encia, 2021.

\bibitem{Kokail:2018eiw}
C.~Kokail {\em et.~al.}, ``{Self-verifying variational quantum simulation of
  lattice models},'' {\em Nature} {\bf 569} (2019), no.~7756 355--360,
  \href{https://arxiv.org/abs/1810.03421}{{\tt 1810.03421}}.

\end{thebibliography}\endgroup

\end{document}